\documentclass[a4paper,12pt]{article}
\pdfoutput=1
\usepackage{epsfig}
\usepackage{amssymb}
\usepackage{amsfonts}
\usepackage{amsmath}
\usepackage{euscript}
\usepackage{verbatim}
\usepackage{latexsym}
\usepackage{graphicx}
\usepackage{caption}
\usepackage{breqn}
\usepackage{float}
\usepackage{subcaption}
\usepackage{placeins}

\usepackage{listings}

\newif\ifdtup

\catcode`\@=11

\@addtoreset{equation}{section}

\def\@normalsize{\@setsize\normalsize{15pt}\xiipt\@xiipt
\abovedisplayskip 14pt plus3pt minus3pt%
\belowdisplayskip \abovedisplayskip
\abovedisplayshortskip \z@ plus3pt%
\belowdisplayshortskip 7pt plus3.5pt minus0pt}

\def\small{\@setsize\small{13.6pt}\xipt\@xipt
\abovedisplayskip 13pt plus3pt minus3pt%
\belowdisplayskip \abovedisplayskip
\abovedisplayshortskip \z@ plus3pt%
\belowdisplayshortskip 7pt plus3.5pt minus0pt
\def\@listi{\parsep 4.5pt plus 2pt minus 1pt
     \itemsep \parsep
     \topsep 9pt plus 3pt minus 3pt}}

\relax

\catcode`@=12

\catcode`\@=11

\def\section{\@startsection{section}{1}{\z@}{3.5ex plus 1ex minus
   .2ex}{2.3ex plus .2ex}{\large\bf}}

\def\SymBoxes#1#2#3#4{\newdimen\un@t \un@t#3%
\raisebox{#1}{\rule{#2\un@t}{#4}\hskip-#2\un@t
\@tempdimb\un@t \advance\@tempdimb by-#4\@tempcntb#2\relax%
\@whilenum{\@tempcntb>0}\do{
\rule{#4}{\un@t}\hskip\@tempdimb \advance\@tempcntb by\m@ne}%
\hskip-#2\un@t \rule[\un@t]{#2\un@t}{#4}%
\rule[\un@t]{#4}{#4}\hskip-#4
\rule{#4}{\un@t}}\hskip-#4}                

\begin{document}

\newcommand{\beq}{\begin{equation}}
\newcommand{\eeq}{\end{equation}}
\newcommand{\bea}{\begin{eqnarray}}
\newcommand{\eea}{\end{eqnarray}}
\newcommand{\beas}{\begin{eqnarray*}}
\newcommand{\eeas}{\end{eqnarray*}}
\newcommand{\defi}{\stackrel{\rm def}{=}}
\newcommand{\non}{\nonumber}
\newcommand{\bquo}{\begin{quote}}
\newcommand{\enqu}{\end{quote}}
\newcommand*{\Resize}[2]{\resizebox{#1}{!}{$#2$}}%
\renewcommand{\(}{\begin{equation}}
\renewcommand{\)}{\end{equation}}
\def \eqn#1#2{\begin{equation}#2\label{#1}\end{equation}}

\def\e{\epsilon}
\def\IZ{{\mathbb Z}}
\def\IR{{\mathbb R}}
\def\IC{{\mathbb C}}
\def\IQ{{\mathbb Q}}
\def\de{\partial}
\def\Tr{ \hbox{\rm Tr}}
\def\H{ \hbox{\rm H}}
\def\HE{ \hbox{$\rm H^{even}$}}
\def\HO{ \hbox{$\rm H^{odd}$}}
\def\K{ \hbox{\rm K}}
\def\Im{ \hbox{\rm Im}}
\def\Ker{ \hbox{\rm Ker}}
\def\const{\hbox {\rm const.}}
\def\o{\over}
\def\im{\hbox{\rm Im}}
\def\re{\hbox{\rm Re}}
\def\bra{\langle}\def\ket{\rangle}
\def\Arg{\hbox {\rm Arg}}
\def\Re{\hbox {\rm Re}}
\def\Im{\hbox {\rm Im}}
\def\exo{\hbox {\rm exp}}
\def\diag{\hbox{\rm diag}}
\def\longvert{{\rule[-2mm]{0.1mm}{7mm}}\,}
\def\a{\alpha}
\def\dag{{}^{\dagger}}
\def\tq{{\widetilde q}}
\def\p{{}^{\prime}}
\def\W{W}
\def\N{{\cal N}}
\def\hsp{,\hspace{.7cm}}

\def\br{\nonumber}
\def\IZ{{\mathbb Z}}
\def\IR{{\mathbb R}}
\def\IC{{\mathbb C}}
\def\IQ{{\mathbb Q}}
\def\IP{{\mathbb P}}
\def \eqn#1#2{\begin{equation}#2\label{#1}\end{equation}}

\newcommand{\C}{\ensuremath{\mathbb C}}
\newcommand{\Z}{\ensuremath{\mathbb Z}}
\newcommand{\R}{\ensuremath{\mathbb R}}
\newcommand{\rp}{\ensuremath{\mathbb {RP}}}
\newcommand{\cp}{\ensuremath{\mathbb {CP}}}
\newcommand{\vac}{\ensuremath{|0\rangle}}
\newcommand{\vact}{\ensuremath{|00\rangle}                    }
\newcommand{\oc}{\ensuremath{\overline{c}}}
\newcommand{\psizero}{\psi_{0}}
\newcommand{\phizero}{\phi_{0}}
\newcommand{\hzero}{h_{0}}
\newcommand{\psiin}{\psi_{\rh}}
\newcommand{\phiin}{\phi_{\rh}}
\newcommand{\hin}{h_{\rh}}
\newcommand{\rh}{r_{h}}
\newcommand{\rb}{r_{b}}
\newcommand{\psibnd}{\psi_{0}^{b}}
\newcommand{\psibndp}{\psi_{1}^{b}}
\newcommand{\phibnd}{\phi_{0}^{b}}
\newcommand{\phibndp}{\phi_{1}^{b}}
\newcommand{\gbnd}{g_{0}^{b}}
\newcommand{\hbnd}{h_{0}^{b}}
\newcommand{\zh}{z_{h}}
\newcommand{\zb}{z_{b}}
\newcommand{\man}{\mathcal{M}}
\newcommand{\hbr}{\bar{h}}
\newcommand{\tbr}{\bar{t}}

\begin{titlepage}
\def\thefootnote{\fnsymbol{footnote}}


\begin{center}
{
{\bf {\Large  What is the Simplest Linear Ramp?} 
}
}
\end{center}
\vspace{-0.2cm}

\begin{center} 
Suman DAS$^a$\footnote{\texttt{suman.das@saha.ac.in}}, Sumit K. GARG$^b$\footnote{\texttt{sumit.kumar@manipal.edu}},
Chethan KRISHNAN$^c$\footnote{\texttt{chethan.krishnan@gmail.com}}, Arnab KUNDU$^a$\footnote{\texttt{arnab.kundu@saha.ac.in}} 

\end{center}

\renewcommand{\thefootnote}{\arabic{footnote}}

\vspace{-0.5cm}

\begin{center}

$^a$ {Theory Division, Saha Institute of Nuclear Physics, \\A CI of Homi Bhabha National Institute,
1/AF, Bidhannagar, Kolkata 700064, India}\\

\vspace{0.1cm}

$^b$ {Manipal Centre for Natural Sciences,
Manipal Academy of Higher Education,\\
Dr. T.M.A. Pai Planetarium Building,
Manipal-576104, Karnataka, India}

$^c$ {Center for High Energy Physics,\\
Indian Institute of Science, Bangalore 560012, India}

\end{center}
\vspace{-0.35cm}
\noindent
\begin{center} {\bf Abstract} \end{center}
\vspace{-0.2cm}
We discuss conditions under which a deterministic sequence of real numbers, interpreted as the set of eigenvalues of a Hamiltonian, can exhibit features usually associated to random matrix spectra. A key diagnostic is the spectral form factor (SFF) -- a linear ramp in the SFF is often viewed as a signature of random matrix behavior. Based on various explicit examples, we observe conditions for linear and power law ramps  to arise in deterministic spectra. We note that a very simple spectrum with a linear ramp is $E_n \sim \log n$. Despite the presence of ramps, these sequences do $not$ exhibit conventional level repulsion, demonstrating that the lore about their concurrence needs refinement. However, when a small noise correction is added to the spectrum, they lead to clear level repulsion as well as the (linear) ramp. We note some remarkable features of logarithmic spectra, apart from their linear ramps:  they are closely related to normal modes of black hole stretched horizons, and their partition function with argument $s=\beta+it$ is the Riemann zeta function $\zeta(s)$. An immediate consequence is that the spectral form factor is simply $\sim |\zeta(it)|^2$. Our observation that log spectra have a linear ramp, is closely related to the Lindel\"of hypothesis on the growth of the zeta function. With elementary numerics, we check that the slope of a best fit line through  $|\zeta(it)|^2$ on a log-log plot is indeed $1$, to the fourth decimal. We also note that truncating the Riemann zeta function sum at a finite integer $N$ causes the would-be-eternal ramp to end on a plateau.

\vspace{1.6 cm}
\vfill

\end{titlepage}

\setcounter{footnote}{0}

\section{Introduction}

In a couple of recent papers \cite{fuzz1, fuzz2}, we considered free scalar field theory on a black hole geometry with a stretched horizon \cite{tHooft}. If one imposes a randomly chosen\footnote{Let us emphasize that what we mean by ``random" is that we choose a {\it single} realization  from an ensemble of profiles (say, suitably Gaussian distributed \cite{fuzz2}), {\em without} any ensemble averages.} profile as the boundary condition for the scalar field at the stretched horizon, this can be viewed as toy  model for a quantum fuzzball and its profile \cite{Mathur, Rychkov, KST, Avinash, superstrata, MathurReview, BenaReview, Martinec}. It was found that despite the simplicity of the set up, the normal modes of this configuration showed three features that one expects from microstates of black holes: 
\begin{itemize} 
\item The spectrum exhibited level repulsion.
\item The Spectral Form Factor (SFF) \cite{Cotler} had a linear ramp of slope $\sim 1$ on a log-log plot. 
\item The linear ramp was not smooth (or averaged) --  it had fluctuations on it. 
\end{itemize} 
The stretched horizon is an ad-hoc construct that needs to be better understood, perhaps from a holographic CFT perspective. But that should be weighed against the fact that there is no other calculation in the literature that we are aware of, that can reproduce more than one of the above three items from the {\em bulk} side of the holographic duality\footnote{See \cite{CVJ} which reproduces the ramp via certain JT gravity calculations motivated by matrix models, but the ramp there is not linear. The (linear) ramp without fluctuations should be reproducible from JT gravity because it is an ensemble average of random matrices \cite{SSS} and therefore should be able to reproduce the sine kernel which is responsible for the ramp \cite{Cotler}. But an inherently ensemble-averaged system does not have fluctuations or level repulsion. JT gravity is a 1+1 dimensional theory -- to contrast, our calculations in \cite{fuzz1,fuzz2} necessarily involve the transverse directions.}. 

The fact that these are {\em free} scalar field theories and yet the spectrum has features reminiscent of random matrices, is striking. In \cite{fuzz2} it was noted that the crucial ingredient in obtaining the linear ramp was the boundary condition {\em close to the horizon}. The choice of the profile was unimportant \cite{fuzz2}, in particular the ramp was intact even when the profile was trivial (i.e., constant) \cite{fuzz1}. This observation is a suggestion that there may exist deterministic spectra that can exhibit the linear ramp, despite the popular expectation that RMT is a key ingredient in the linearity of the ramp. To contrast, level repulsion arises \cite{fuzz2} because of the small noise-like fluctuations in the spectrum that arise as a consequence of the variations in the boundary condition profile. This is an indication that the linear ramp captures the long-range (and therefore more robust) features of the spectrum whereas level repulsion is related to its short-range behavior. 

Motivated by these observations, in this paper, we will study the question of the linear ramp without any direct reference to black holes. Our central question is: are there deterministic sequences of real numbers, interpreted as the spectrum of a Hamiltonian arranged monotonically, that can give rise to a (linear) ramp? We will present strong evidence that the answer to this question is ``yes". A summary of key results as well as some discussion, is presented in the final section.

In this short note, which is a follow-up of the works in \cite{fuzz1, fuzz2}, we will not review basic definitions of quantities like SFF and level spacing distribution. The reader is referred to these previous papers (as well as standard references therein) for those definitions, here we will only discuss the new results. We expect however that the discussion below should be quite accessible to those with a nodding acquaintance of the fact that ``spectral form factors of random matrices have a dip-ramp-plateau structure, with a linear ramp".

\section{Deterministic Ramps}

Our goal is to get some intuition about the genericity of ramps in spectra where we know the eigenvalue $E_n$ as a deterministic function of $n$. Of course, if we pick a Hamiltonian from an RMT ensemble and consider its eigenvalues, it is clear that we will get a ramp. But this procedure is not deterministic in the sense we use it, because it involves sampling from an RMT ensemble. 

We will start our discussion by considering spectra of the form 
\bea
E_n \sim n^\alpha
\eea
where $\alpha$ is a real parameter that we tune. The overall scale  of $E_n$ will not affect the behavior of the SFF, and can be viewed as determining the unit of time. We will typically set it to 1 in our calculations. For $\alpha \ge 1 $  these systems do not exhibit any clear ramp structure, but by numerical experiments we find that a ramp (whose slope is fixed for a given $\alpha$) begins to emerge when $\alpha < 1$. Let us emphasize here that when we discuss the slope of a ramp, we always mean its slope in a log-log plot. Following convention, a ramp of slope 1 in a log-log plot will be referred to as a linear ramp -- if the slope is constant but different from 1, we will call it a power law ramp. For generic values of $\alpha < 1$, we find a power law ramp.

The values $\alpha =1$ and $\alpha =0$ are special in some ways. The former corresponds to the simple harmonic oscillator (note that constant shifts in the spectrum do not affect the SFF). This is the ultimate integrable system, and there is no ramp\footnote{We will see later that the ramp appears even in the SHO spectrum, when you add a small amount of noise to the spectrum. But this is moving away from the deterministic requirement, so we will discuss it in a later section.} (linear or otherwise) at and above this threshold value of $\alpha$. We start seeing a power law ramp (i.e., ramp of constant slope $\neq 1$) for values of $\alpha <1$. The slope of the ramp steadily decreases as we decrease $\alpha$ from 1 to zero, reaching the sought for linear ramp (i.e., slope 1), in the limit $\alpha \rightarrow 0$. If we decrease the value of $\alpha$ further down below zero, we find that the slope of the ramp steadily decreases below 1. Our numerics are loosely consistent with the possibility that the slope of the ramp for the spectrum $E_n \sim n^\alpha$ is $\sim 1/(1-\alpha)$, it may be interesting to establish a more precise statement about the slope. Our focus in this paper will be on the linear ramp of slope 1, and therefore on the $\alpha \rightarrow 0$ limit.
\begin{figure}[H]
     \centering
     \begin{subfigure}[b]{0.3\textwidth}
         \centering
         \includegraphics[width=\textwidth]{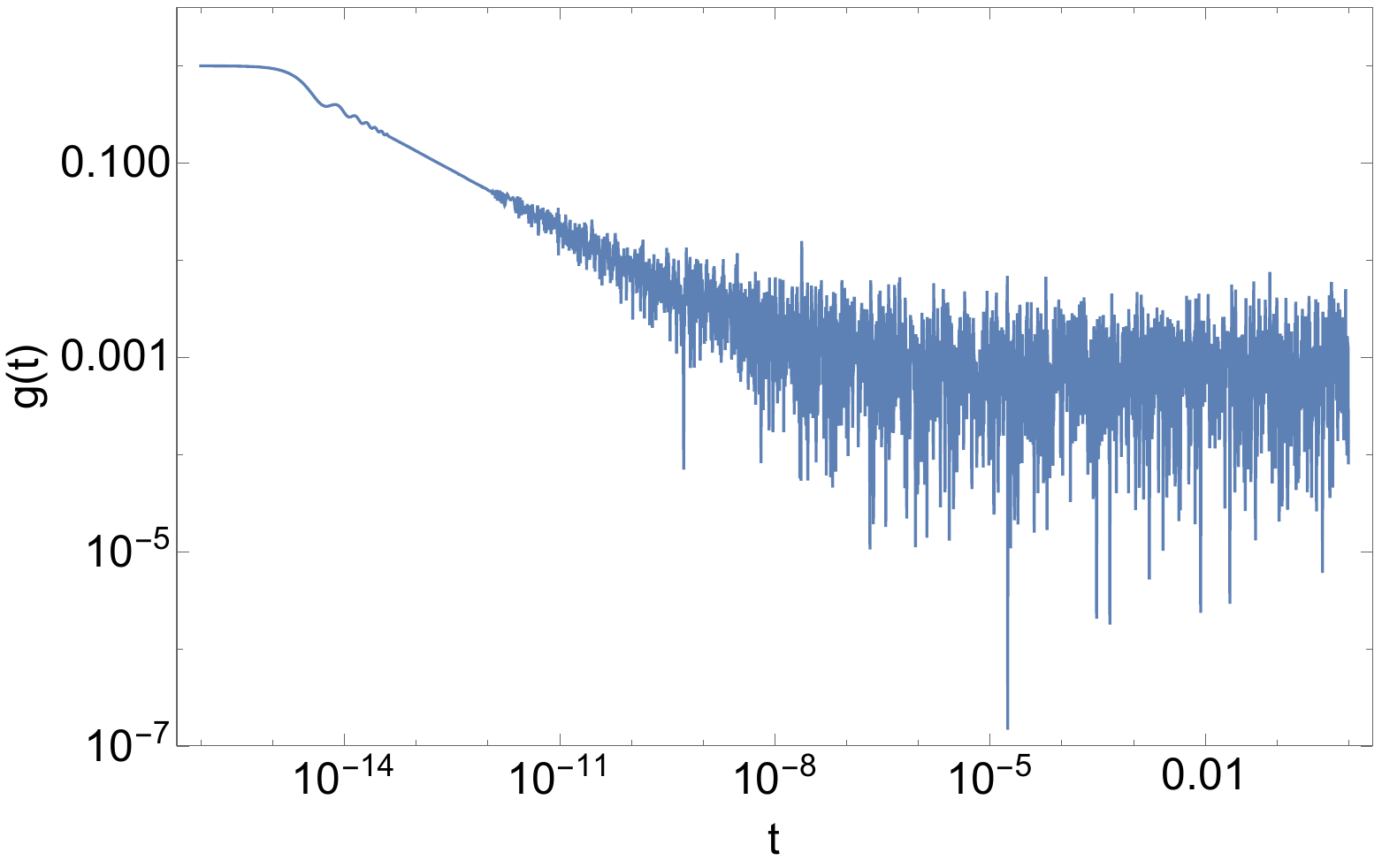}
         \caption{$\alpha=5.0$}
         \label{fig_1_1}
     \end{subfigure}
     \hfill
      \begin{subfigure}[b]{0.3\textwidth}
         \centering
         \includegraphics[width=\textwidth]{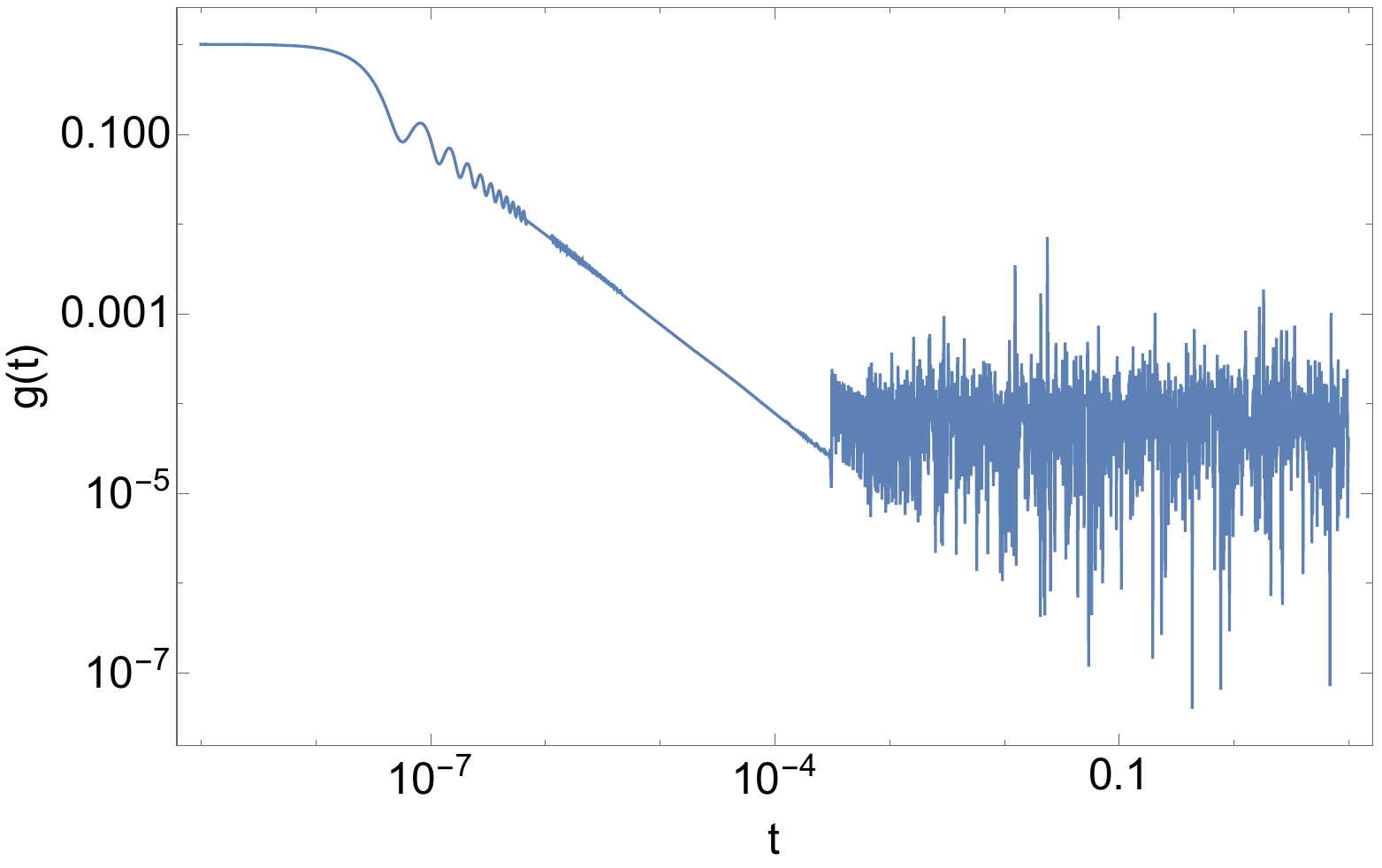}
         \caption{$\alpha=2.0$}
         \label{fig_1_05}
     \end{subfigure}
     \hfill
     \begin{subfigure}[b]{0.3\textwidth}
         \centering
         \includegraphics[width=\textwidth]{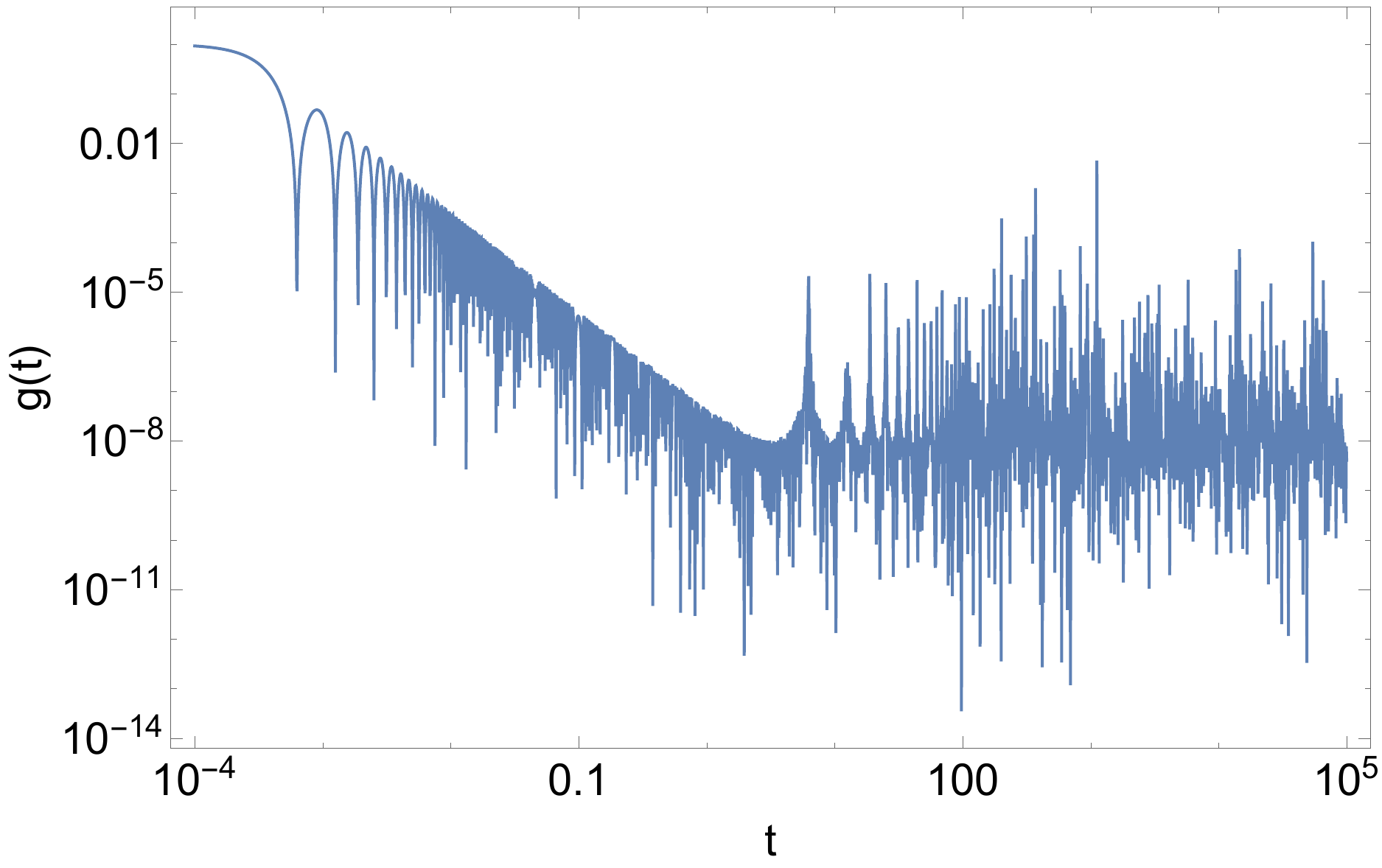}
         \caption{$\alpha=1.0$}
         \label{fig_1_005}
     \end{subfigure}
     \hfill
  \begin{subfigure}[b]{0.3\textwidth}
         \centering
         \includegraphics[width=\textwidth]{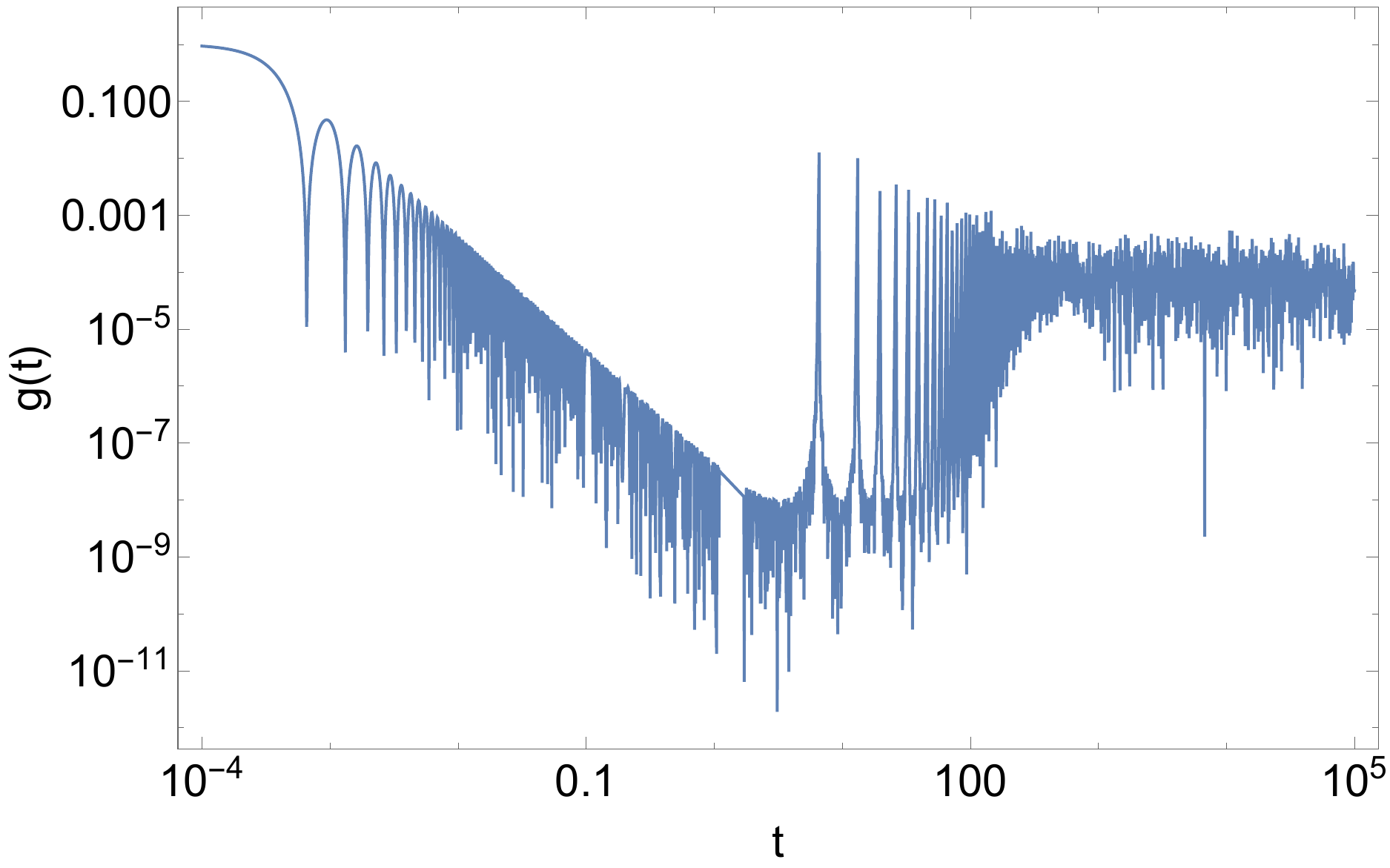}
         \caption{$\alpha=1/1.005$}
         \label{fig_1_005}
     \end{subfigure}
     \hfill
     \begin{subfigure}[b]{0.3\textwidth}
         \centering
         \includegraphics[width=\textwidth]{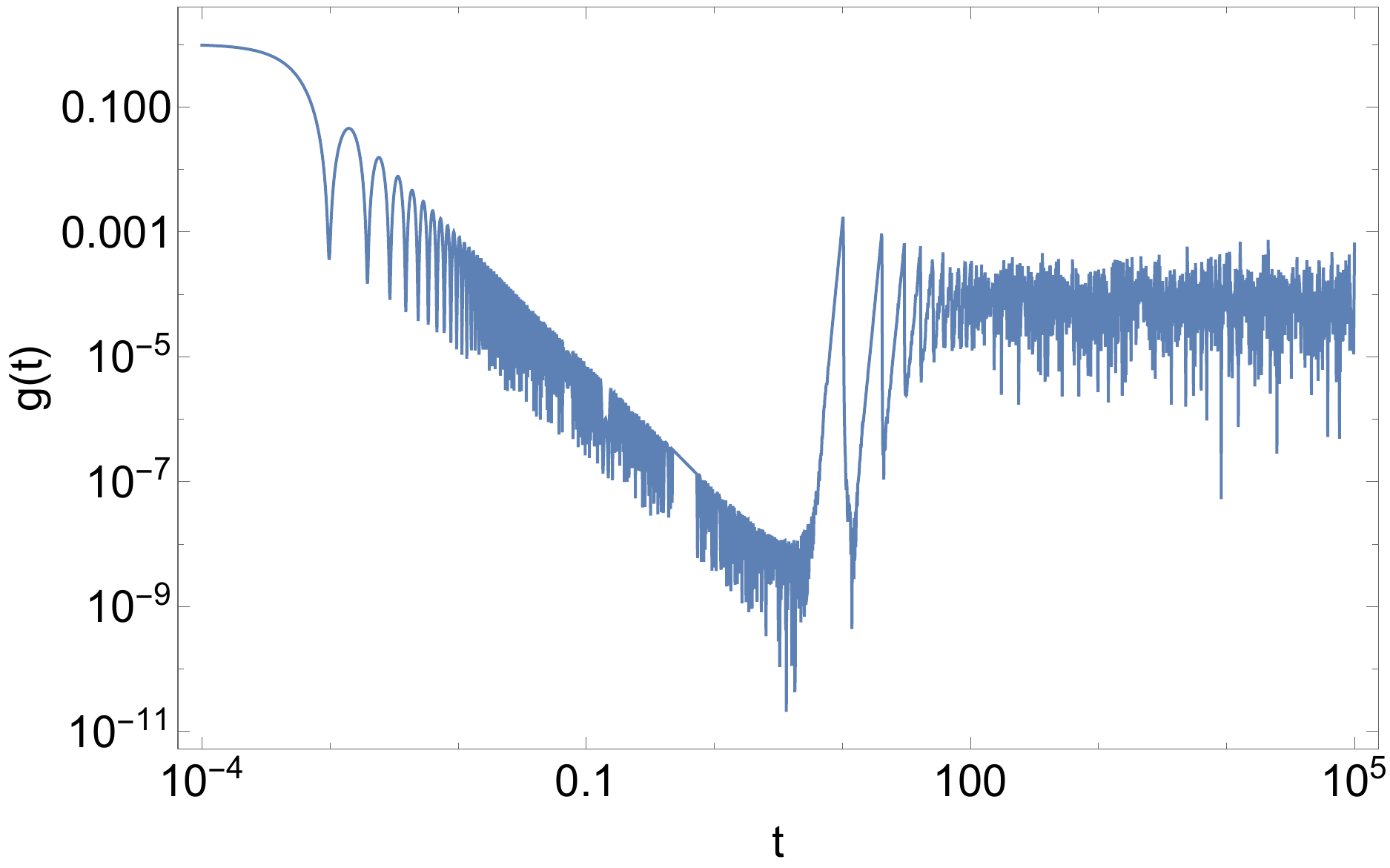}
         \caption{$\alpha=1/1.05$}
         \label{fig_1_05}
     \end{subfigure}
     \hfill
     \begin{subfigure}[b]{0.3\textwidth}
         \centering
         \includegraphics[width=\textwidth]{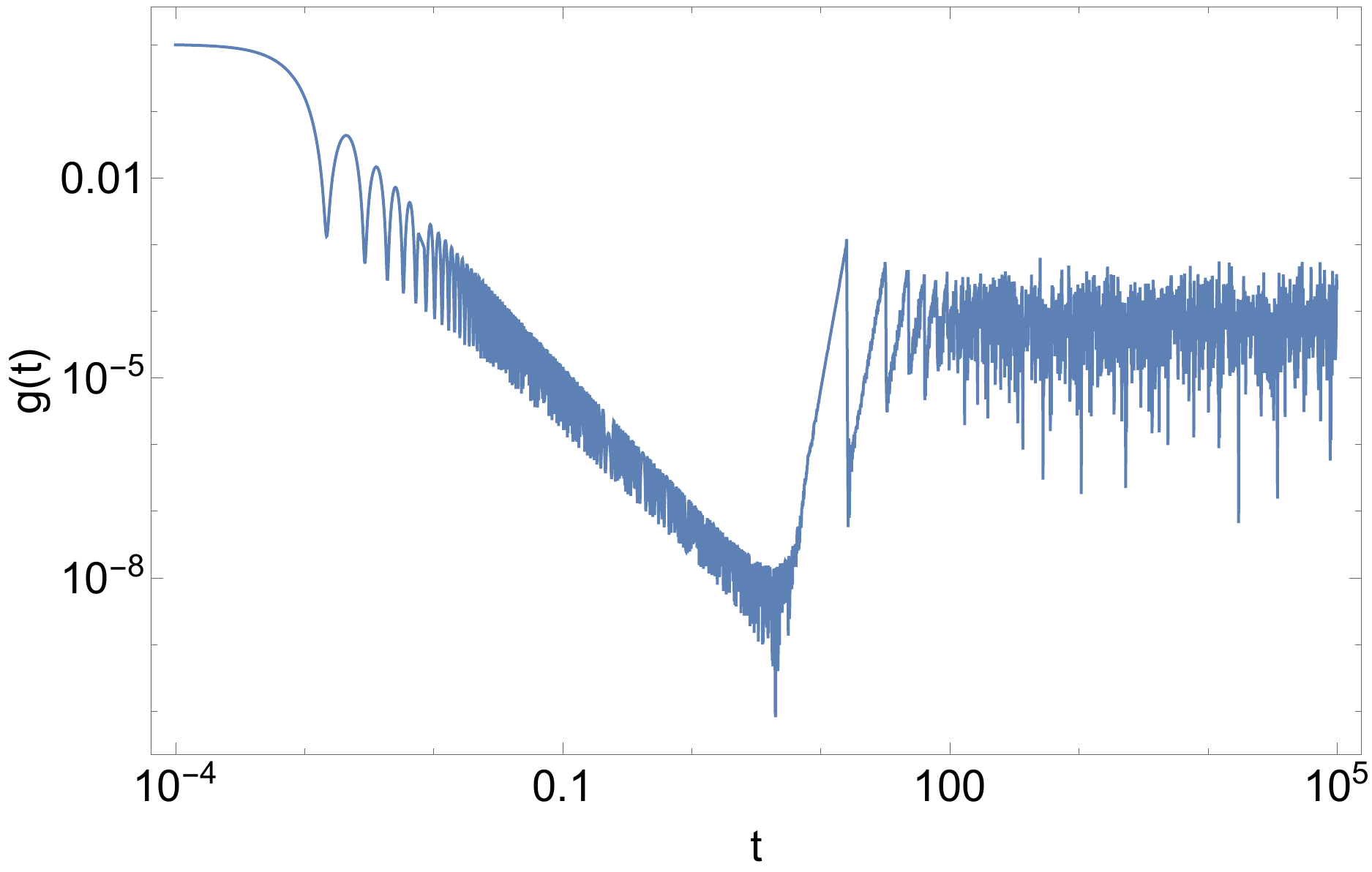}
         \caption{$\alpha=1/1.1$}
         \label{fig_1_1}
     \end{subfigure}
      \hfill
     \begin{subfigure}[b]{0.3\textwidth}
         \centering
         \includegraphics[width=\textwidth]{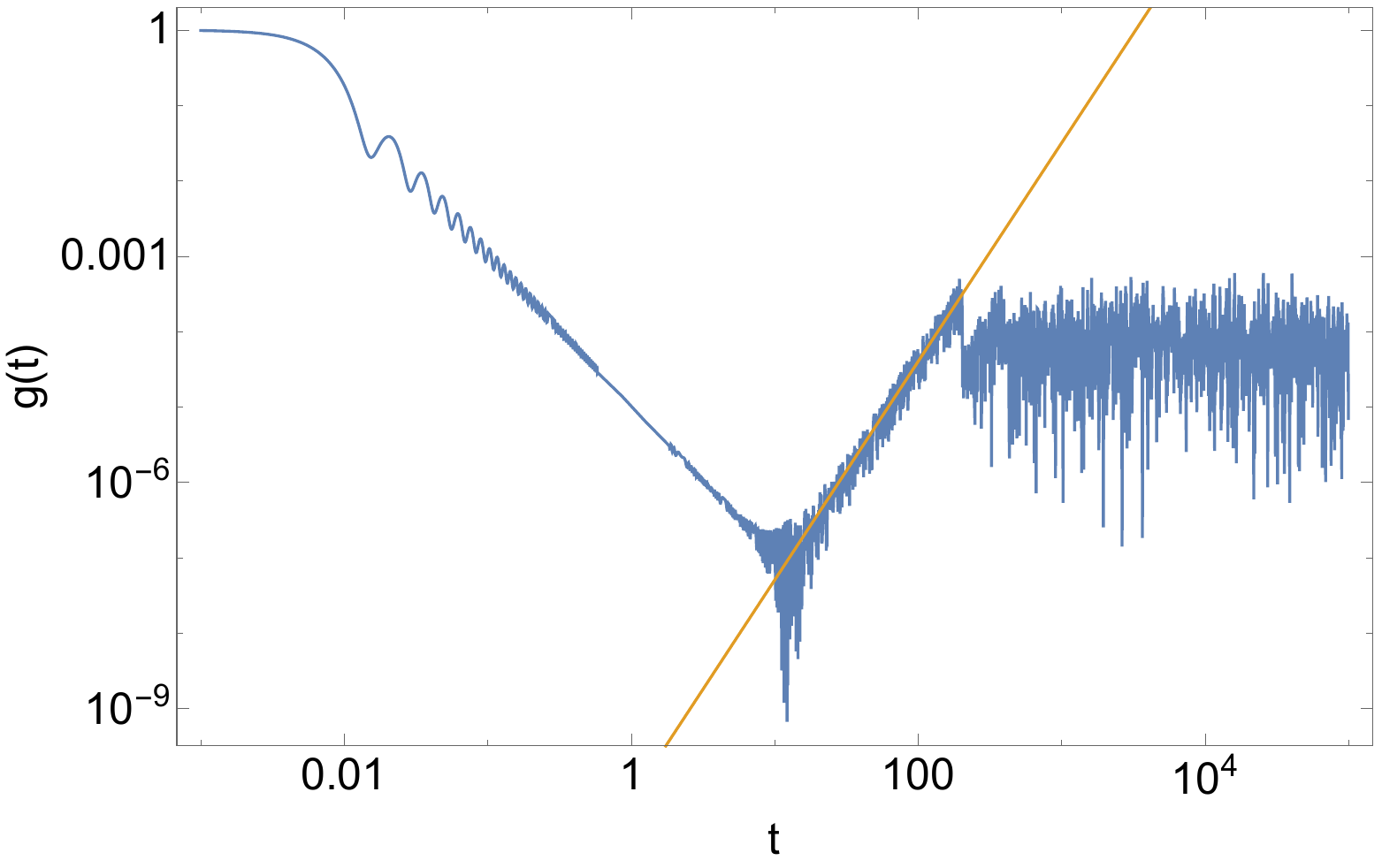}
         \caption{$\alpha=1/1.5$, slope $\sim 2.9$}
         \label{fig_1_5}
     \end{subfigure}
      \hfill
     \begin{subfigure}[b]{0.3\textwidth}
         \centering
         \includegraphics[width=\textwidth]{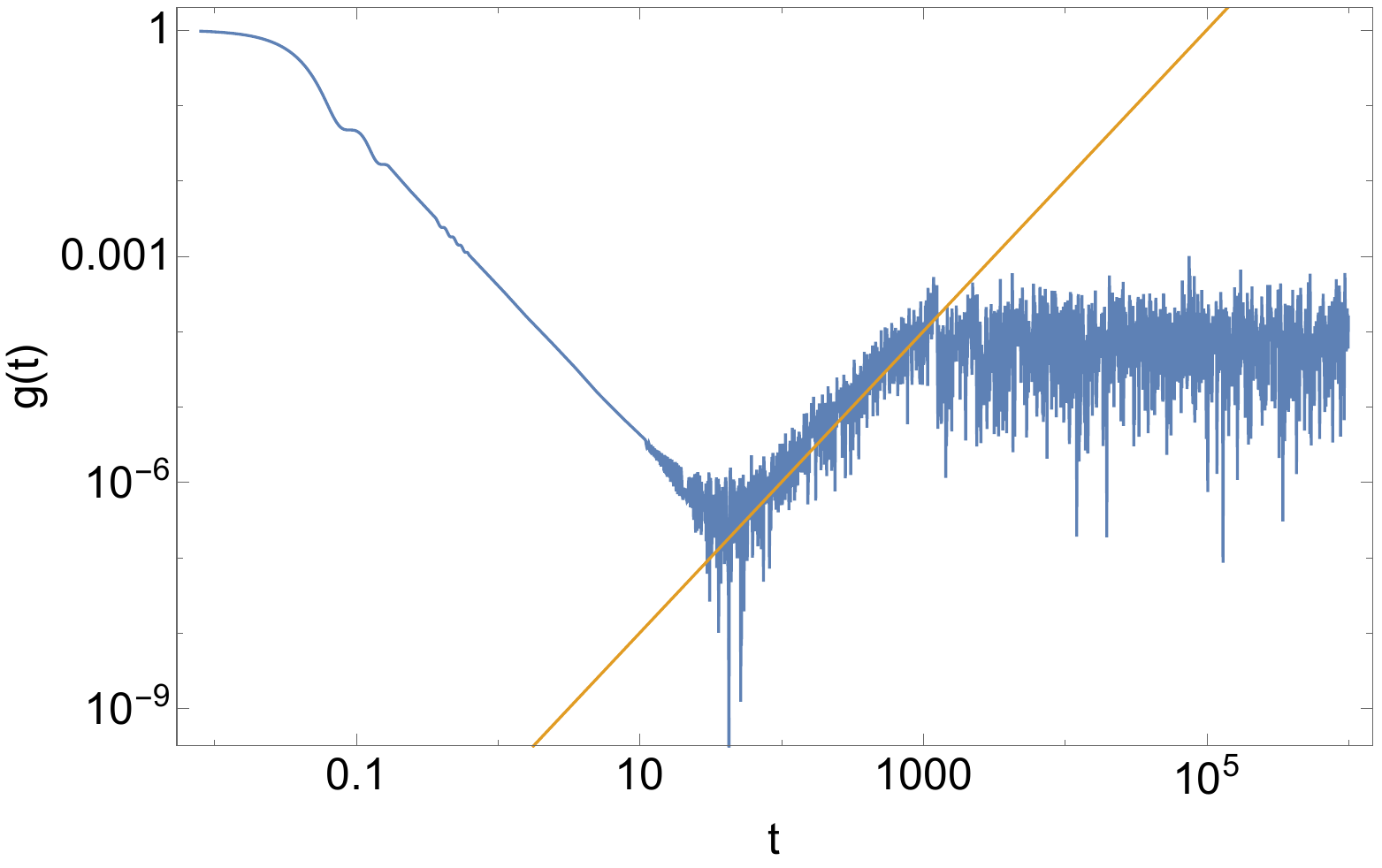}
         \caption{$\alpha=1/2$, slope $\sim 2$}
         \label{fig_2}
     \end{subfigure}
      \hfill
     \begin{subfigure}[b]{0.3\textwidth}
         \centering
         \includegraphics[width=\textwidth]{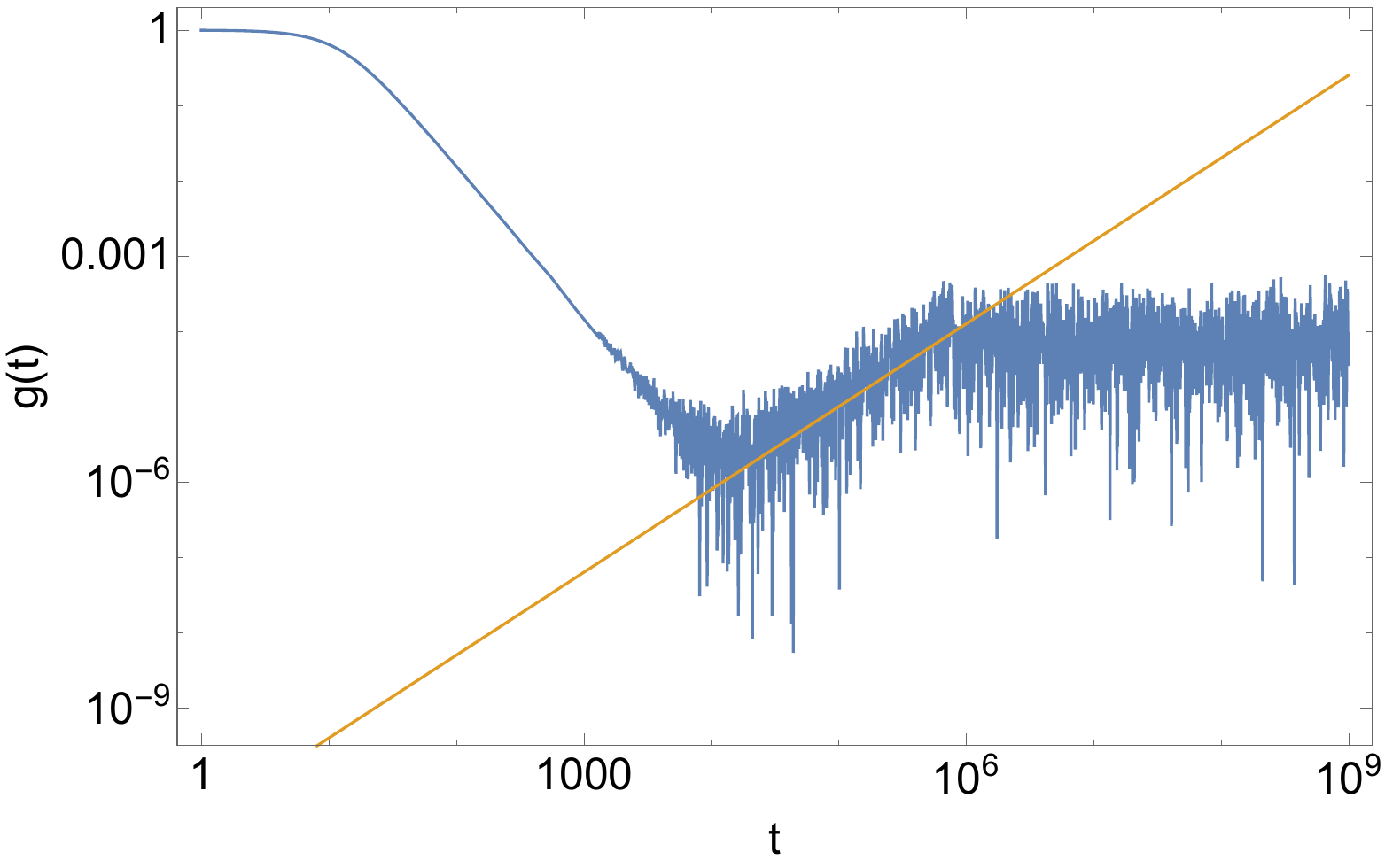}
         \caption{$\alpha=1/20$, slope $\sim 1.1$}
         \label{fig_20}
     \end{subfigure}
     \hfill
     \begin{subfigure}[b]{0.3\textwidth}
         \centering
         \includegraphics[width=\textwidth]{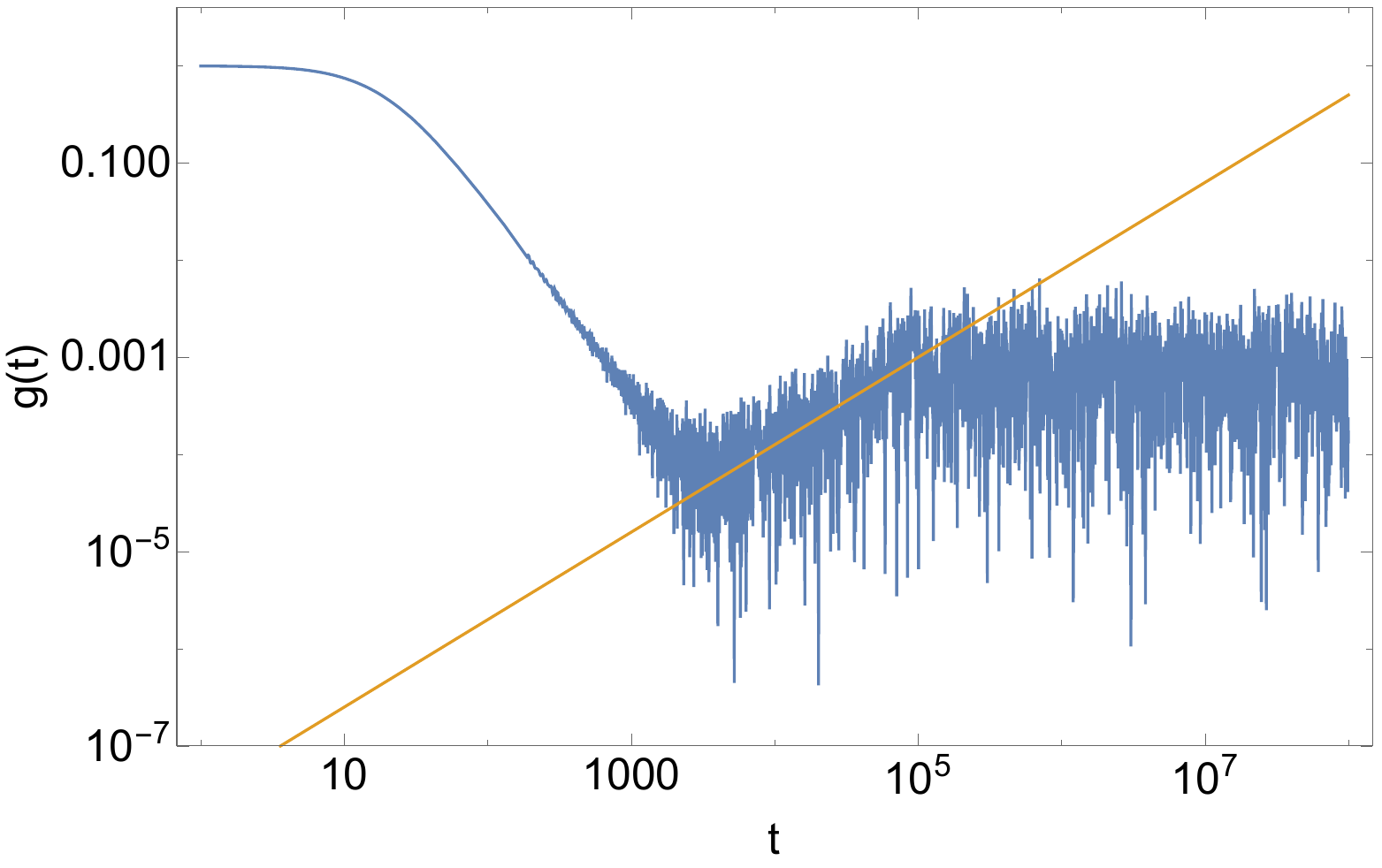}
         \caption{$\alpha=-0.1$, slope $\sim 0.9$}
         \label{fig_1_005}
     \end{subfigure}
     \hfill
     \begin{subfigure}[b]{0.3\textwidth}
         \centering
         \includegraphics[width=\textwidth]{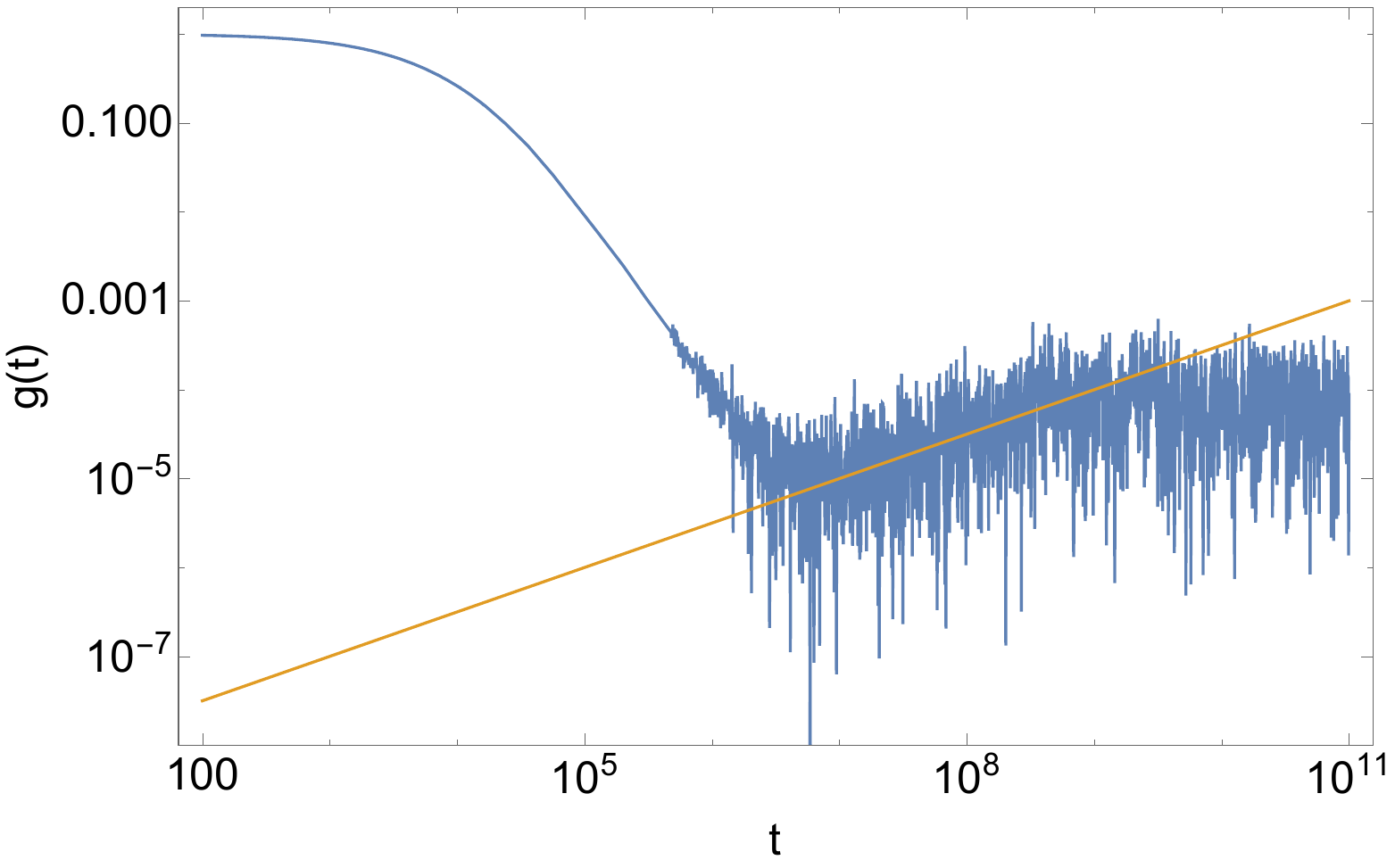}
         \caption{$\alpha=-1$, slope $\sim 0.5$}
         \label{fig_1_05}
     \end{subfigure}
     \hfill
     \begin{subfigure}[b]{0.3\textwidth}
         \centering
         \includegraphics[width=\textwidth]{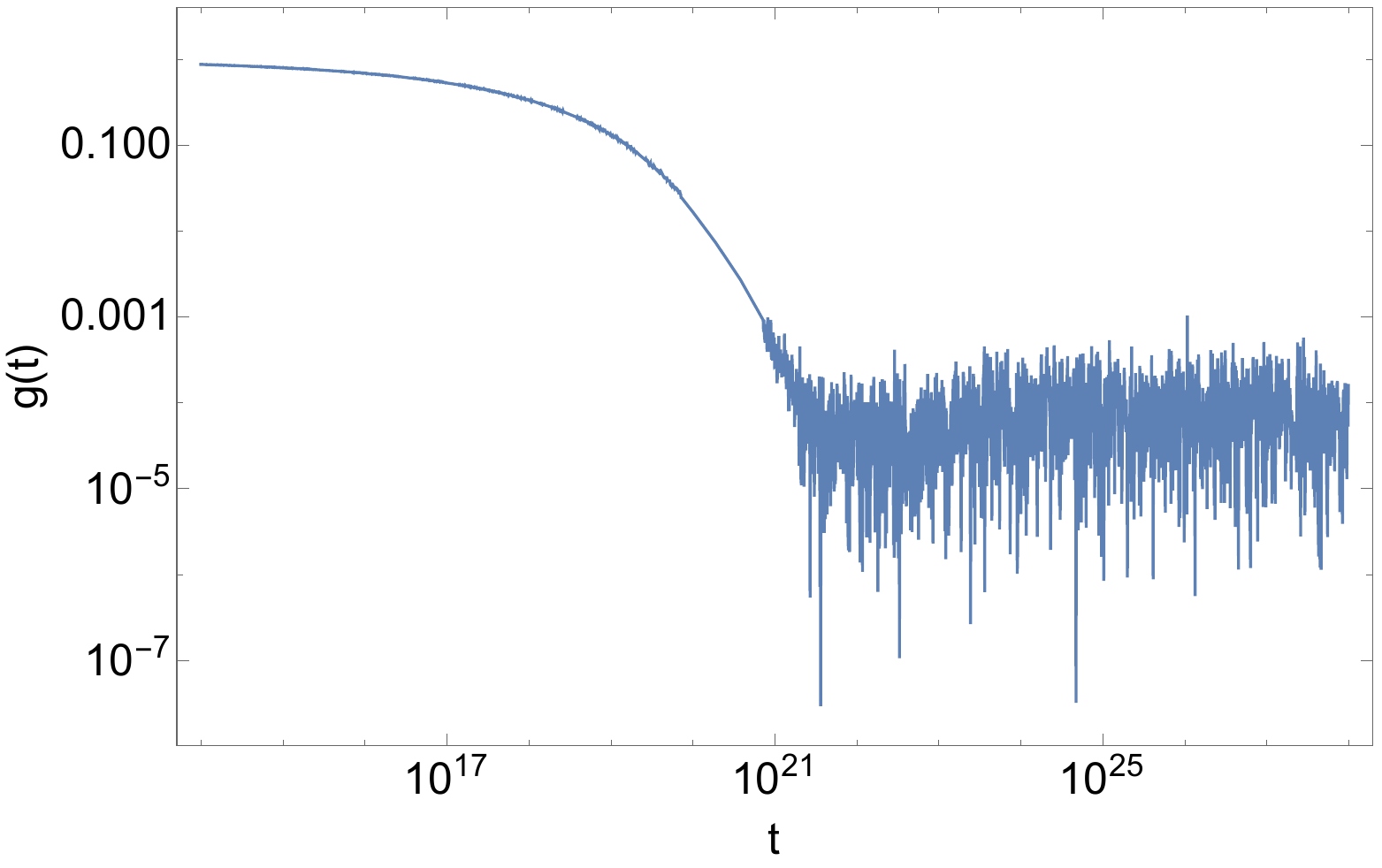}
         \caption{$\alpha=-5.0$}
         \label{fig_1_1}
     \end{subfigure}
        \caption{SFF for power law spectra $E_n=\{n^{\alpha}\}$ with $n_{cut}=10000$ and $\beta=0$. Slope of yellow straight lines are written in the captions. A ramp-structure starts appearing below $\alpha =1$, becoming increasingly cleaner as $\alpha$ decreases.}
        \label{n_alpha_sff}
\end{figure}

Of course, $\alpha =0$ corresponds to a spectrum that is degenerate, and it is not very meaningful to say that it has a linear ramp. What we really mean by $\alpha \rightarrow 0$ limit is that we can progressively consider smaller and smaller values of $|\alpha| \rightarrow 0$, and compute the SFF for each of these values -- for any sufficiently small  value of $|\alpha|$ the slope of the ramp is numerically consistent with 1.  A closely related idea (as we will argue below) is to consider the log spectrum
\bea
E_n \sim \log n.
\eea
One can heuristically view this as arising from  $n^\alpha = e^{\alpha \log n} \sim 1+\alpha \log n$ in the small $\alpha$ limit, because shifts and re-scalings of the spectrum do not affect the ramp. It is easy to check that the log spectrum has a linear ramp directly, and we will discuss it in more detail in the next section.

Before we delve in detail into the linear ramp and the log spectrum, let us make a few more general comments. We have found via numerical experimentation, that a key ingredient needed for the presence of a ramp (linear or power law) in many sequences is that 
\bea
\lim_{n \rightarrow \infty} \log_n E_n < 1. \label{conj}
\eea
Note that the base of the logarithm is $n$.
Our comments above about the $E_n \sim n^\alpha$ spectrum can be viewed as a more concrete instantiation of \eqref{conj}. Together with the fact that we can arrange spectra in monotonically non-decreasing sequences which are bounded from below and can use constant shifts to make the ground state energy non-negative\footnote{In particular, all the $E_n$'s we consider are positive, or at least non-negative.}, condition \eqref{conj} is a fairly general criterion. We present some examples of sequences which do or do not have power law ramps in Table \ref{table1}, which are in accord with this criterion. Our goal here is not to offer an exhaustive necessity/sufficiency condition on when ramps exist, but to emphasize that they are not hard to arrange in deterministic sequences. 
Crowding in the levels as a function of $n$ is often what decides the existence and the slope of the ramp. There is a lot of structure in this quantity, and it may be worth exploring this in more detail. Our primary work horse in this paper is $E_n \sim \log n$ which satisfies criterion \eqref{conj} and in fact has a linear ramp, as can be checked. While it may be of interest to study more exotic looking sequences, our focus in this paper is on simplicity and physical and mathematical relevance.
\begin{table}[h]
\centering
\begin{tabular}{|c|c|c|}
 \hline 
 Ramp & $\log n, 2^{1/n}, \pi^{1/n}, n^{1/n}, n^{1/n^2}, n^{2/n}, e^{1/n}, n^\alpha\ \text{with}\hspace{0.6mm} \alpha<1$, ... \\ 
 \hline
 No Ramp & $n^{\log n}, (\log n)^n, 
 n^n, e^n, n^{\alpha}\ \text{with}\hspace{0.6mm} \alpha>1$, ...  \\
 \hline
\end{tabular}
\caption{Some sequences with and without power law ramps. The existence and the (reasonable) ease with  which ramp-y  sequences can be found, is the main point here. The inverse problem of constructing Hamiltonians with these spectra is a problem we do not address systematically.}
\label{table1}
\end{table}

\section{God's Own Linear Ramp}

\begin{figure}[H]
     \centering
     \begin{subfigure}[b]{0.3\textwidth}
         \centering
         \includegraphics[width=\textwidth]{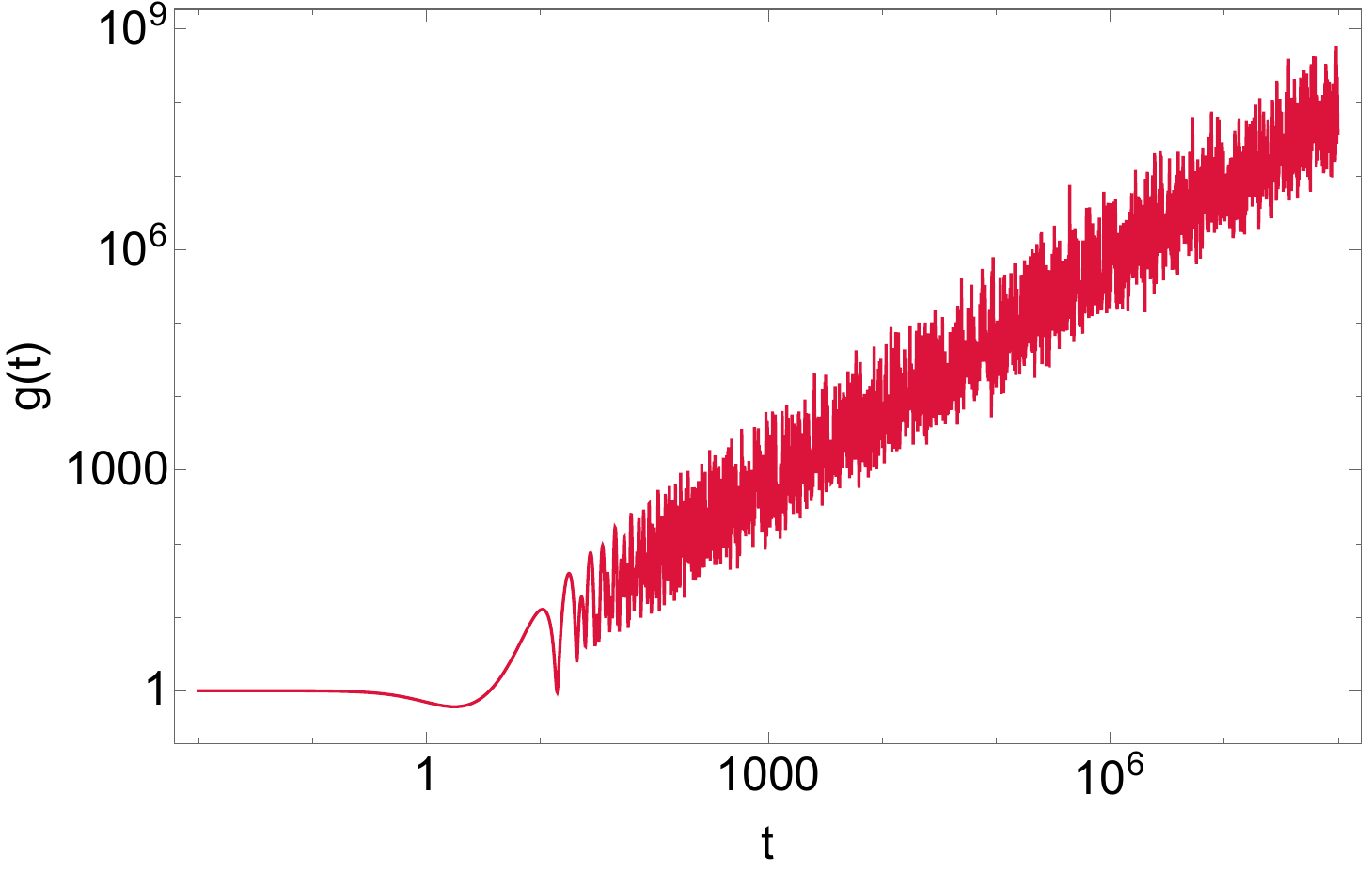}
         \caption{}
         \label{figg1}
     \end{subfigure}
     \hfill
     \begin{subfigure}[b]{0.3\textwidth}
         \centering
         \includegraphics[width=\textwidth]{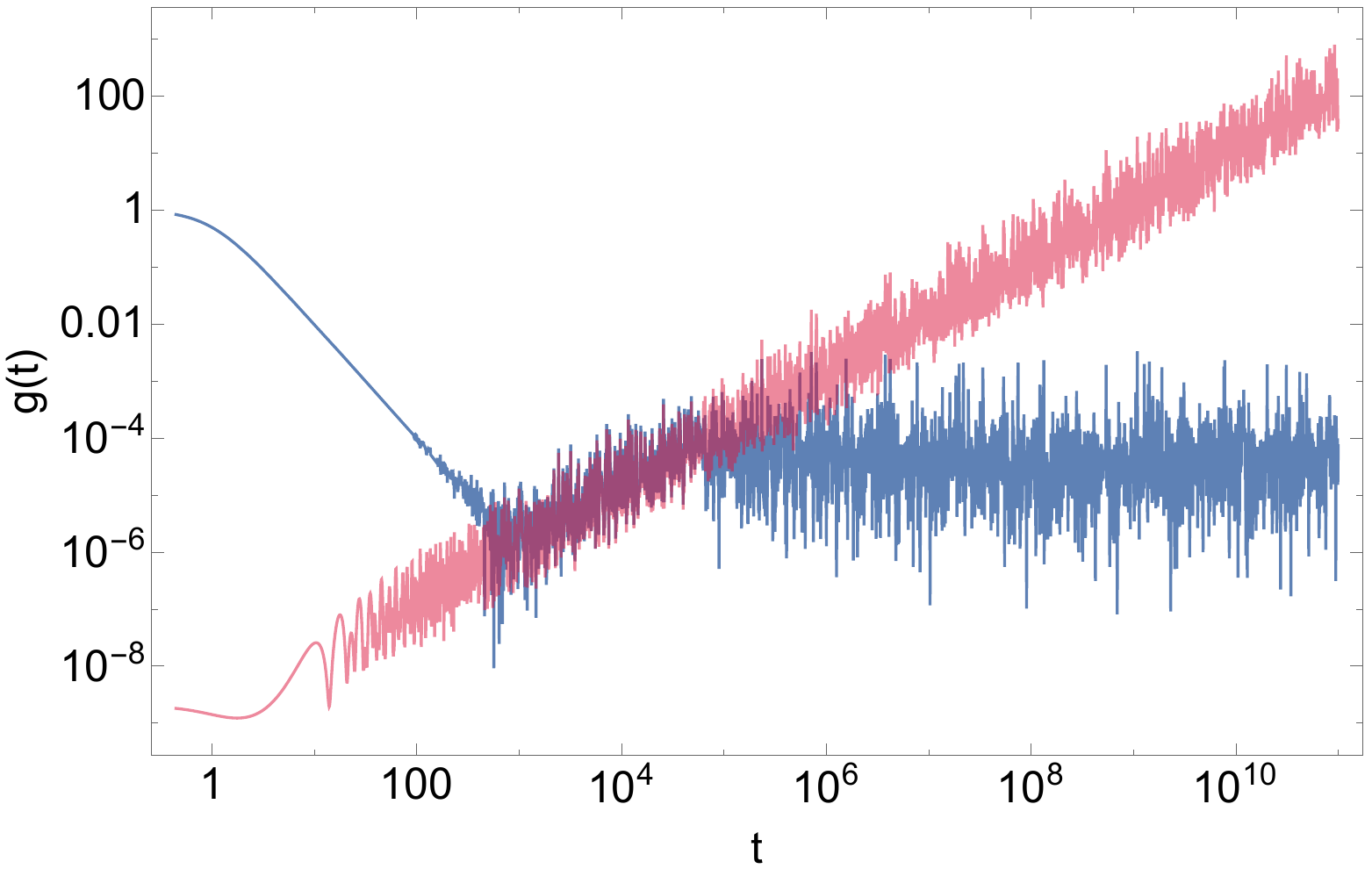}
         \caption{}
         \label{figg2}
     \end{subfigure}
     \hfill
     \begin{subfigure}[b]{0.3\textwidth}
         \centering
         \includegraphics[width=\textwidth]{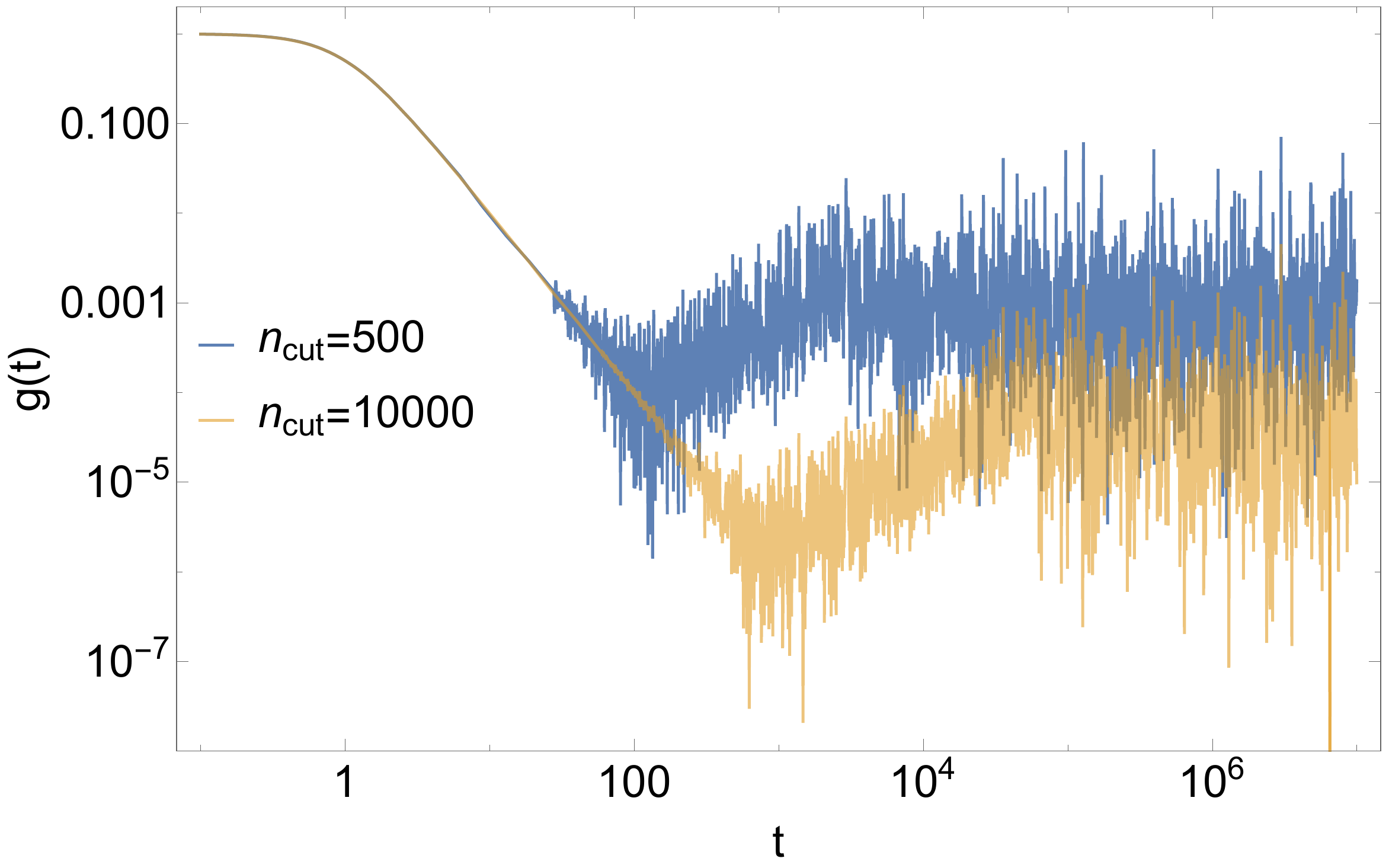}
         \caption{}
         \label{figg3}
     \end{subfigure}
        \caption{(a) SFF of log spectrum without truncation at which gives rise to an eternal ramp. (b) SFF of log spectrum with (blue) and without (red) truncation at $n_{cut}$. Truncation produces plateau in the SFF. (c) Behavior of SFF of truncated log spectrum as we vary $n_{cut}$. }
        \label{zeta_SFF}
\end{figure}

A trivial consequence of working with the log spectrum $E_n=\log n$ is that the partition function of the system at the complexified temperature $s \equiv \beta + i t$ is the Riemann zeta function $\zeta(s)$. It is only when $\beta >1$ that the familiar sum over integers definition of the zeta function (which is what our log spectrum leads to) converges. But often in the RMT literature (see e.g. \cite{Altland}) one is interested in the $\beta=0$ case. There are two ways one can address this problem. One is to truncate the sum at some finite integer $n=n_{cut}$, as maybe expected on physical grounds in general systems with a finite number of degrees of freedom. We have investigated this as well (see e.g., Figure \ref{zeta_SFF}), but we will first proceed via a different route.
 
The key observation is that since the zeta function is a meromorphic function on the complex plane with a pole only at $s=1$, it can be defined by analytic continuation at $\beta=0$ as well\footnote{That such an analytic continuation exists was indeed the opening result of Riemann's illustrious paper.}. This leads to the standard normalized SFF at $\beta=0$ through
\bea
 g(t) = \frac{|\zeta(it)|^2}{|\zeta(0)|^2}, \label{RiemannSFF}
\eea\label{zeta_SFF_eq}
where it is understood that the zeta function is defined via analytic continuation. Plotting this SFF (see Figure \ref{zeta_SFF}), we immediately see that it leads to a very clear ramp. 
In fact, a second and equally immediate observation is that the ramp is ``eternal" and that it never ends at a plateau\footnote{Because it has the looks of a ramp to the heavens, we will call it the heavenly (or eternal or Platonic) ramp.}. This we will view as a consequence of the fact that we have defined the SFF {\em without} truncating the spectrum at some $n_{cut}$. A well-known fact is that the plateau is a non-perturbative effect in RMT systems, and our result is morally analogous to that.  The ramp on the other hand is supposed to be a perturbative effect in the random matrix size \cite{Cotler}. In other words, we should be able to see the ramp at infinite $n_{cut}$ (by doing perturbation theory around it). This can be regarded as explaining why we are able to see the ramp in the (analytically continued) RZF without any truncation at $n_{cut}$, while to see the plateau we need the finite $n_{cut}$ truncation\footnote{An aside worth noting here however is that when comparing random matrix systems with typical large-$N$ theories, the matrix size corresponds to Hilbert space size and therefore scales as $e^N$ in the gauge group rank. In other words $n_{cut} \sim e^N$. So the ramp is non-perturbative and plateau is doubly non-perturbative, in this language.}. One can check that the slopes of the eternal and truncated-in-$n_{cut}$ ramps are mutually consistent, see Figure \ref{zeta_SFF}. 

\begin{figure}[H]
    \centering
    \includegraphics[width=0.75\textwidth]{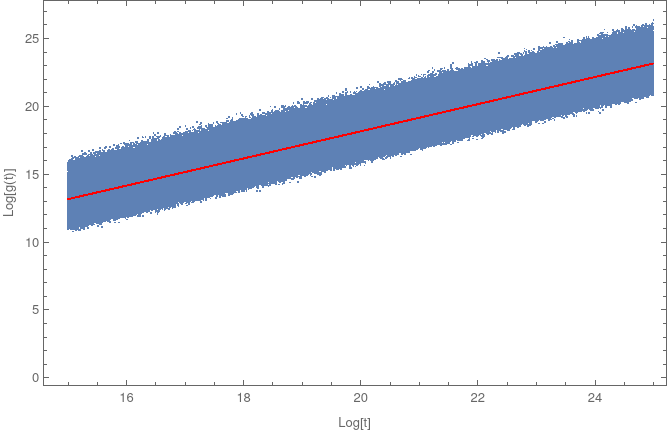}
    \caption{The eternal ramp of the RZF allows us the possibility of looking for a best fit line. In this plot, we consider $g(t)$ at $10^6$ equally spaced (in $\log t$) points between $t=10^{15}$ and $t=10^{25}$ and find the best fit line through these points. The result is the red line:  $\log g(t)=-1.84085 + 1.00017 \log t$.}
    \label{RRamp}
\end{figure}

The advantage of an analytic expression of the form \eqref{RiemannSFF} for the SFF with a seemingly infinite ramp is that we can evaluate the slope numerically to fairly high precision by simply finding the best fit line through a large (temporal) sample of points. When the ramp is of finite length (with edge effects near the dip and the plateau) and the number of eigenvalues is finite, this is a less well-defined procedure. Often one is limited to plotting a slope 1 line, adjusting the intercept and eyeballing to see if the fit looks reasonable\footnote{This is not a particularly serious problem in many cases, because one's goals are often qualitative, and a well-defined ramp with a slope even close to 1 does not happen in integrable systems.}. The Platonic ramp of the Riemann zeta SFF provides us a way beyond this. Quite effortlessly, we are able to find best fit slope 1 lines that are valid to the fourth decimal, see Figure \ref{RRamp}. 

There are three natural quantities that control the precision/accuracy of this slope, and they are (a) the length of the time sample that we consider, (b) the number of points we sample, and (c) the measure that we use in $t$ for the sampling. However, it is not very clear to us how these features affect the precision of our slope -- the trouble is that because the slope is on a log-log plot, it is challenging to go to sufficiently higher values of $t$ so that there is a discernible improvement in the slope. In any event, it is remarkable that even without much numerical sophistication (and without the ability to reach much higher values of time), we find a slope that is very closely consistent with 1. In fact, this is one of the best linear ramps that we have seen in numerical discussions of SFF in SYK and related systems. 

It will be interesting to see if the slope can be argued to average to 1 as we increase the time window we consider, using some analytic argument. Our above observation on the slope is in fact very closely related to a known conjecture about the Riemann zeta function, called the Lindel\"of hypothesis. Lindel\"of bounds are bounds on the growth of the Riemann zeta function characterized by the function $\mu(\beta)$ which is the least upper bound of the numbers $A$ such that $|\zeta(\beta + it)|\ t^{-A}$  is bounded as $t \rightarrow \infty$. For $\beta=0$, which is our case of interest, it is known \cite{Edwards} that $\mu(\beta)=\frac{1}{2}$. A sufficient (but likely not necessary) condition for our claim about the linearity of the ramp to be satisfied is that for large values of $t$ the bound is ``saturated" in the sense that 
\bea
\lim_{t \rightarrow \infty} \frac{\log |\zeta(it)|}{\log t}   = \frac{1}{2}.
\eea
From the fit in Figure \ref{RRamp}, it should be clear that numerically this holds quite well.
It is noteworthy that the Lindel\"of bound at $\beta=0$ turns out to be precisely that value that allows a non-trivial statement of this type. If $\mu(0)$ was $ > 1/2$ there would have been no interesting statement to relate to regarding the linear ramp, while if $\mu(0)$ was $< 0$ it would rule out the linearity of the ramp. Remarkably, neither of these is true. The Lindel\"of bound at $\beta=0$ is the value of the bound that would allow a linear ramp at (asymptotic) saturation.  
It may also be worth  pointing out that while the bound for $\beta=0$ is a proven theorem, on the critical line ($\beta=1/2$) it is still a conjecture. In fact it is a conjecture that is known to be implied by the Riemann hypothesis \cite{Wiki}. Clearly our observation about the slope of the SFF is closely related to the Lindel\"of hypothesis, but it is more of an ``averaged" version of the statement applying at large $t$ at the $\beta=0$ boundary of the critical strip.

Let us make one more comment. The critical zeroes of the Riemann zeta function have been studied intensely \cite{Odlyzko}. The pair correlation function of the zeroes has been shown numerically to have the same form as that of a random matrix. This means that the if we were able to do a Fourier transform of this quantity (which defines an SFF of the zeroes \cite{Cotler}), we would {\em expect} to see a linear ramp. Note however that this is not the object we are studying. We are  not computing the SFF of the zeroes, instead we are led to view the RZF itself (more precisely, its squared modulus) as the SFF. In other words, the spectrum here is $\log n$, not the zeroes of Riemann. So we are not aware of a simple way in which the linear ramp here is connected to the RMT behavior of Riemann zeroes.

\section{The Uses of Noisy Spectra} 

\begin{figure}[H]
\begin{subfigure}{0.47\textwidth}
    \centering
    \includegraphics[width=\textwidth]{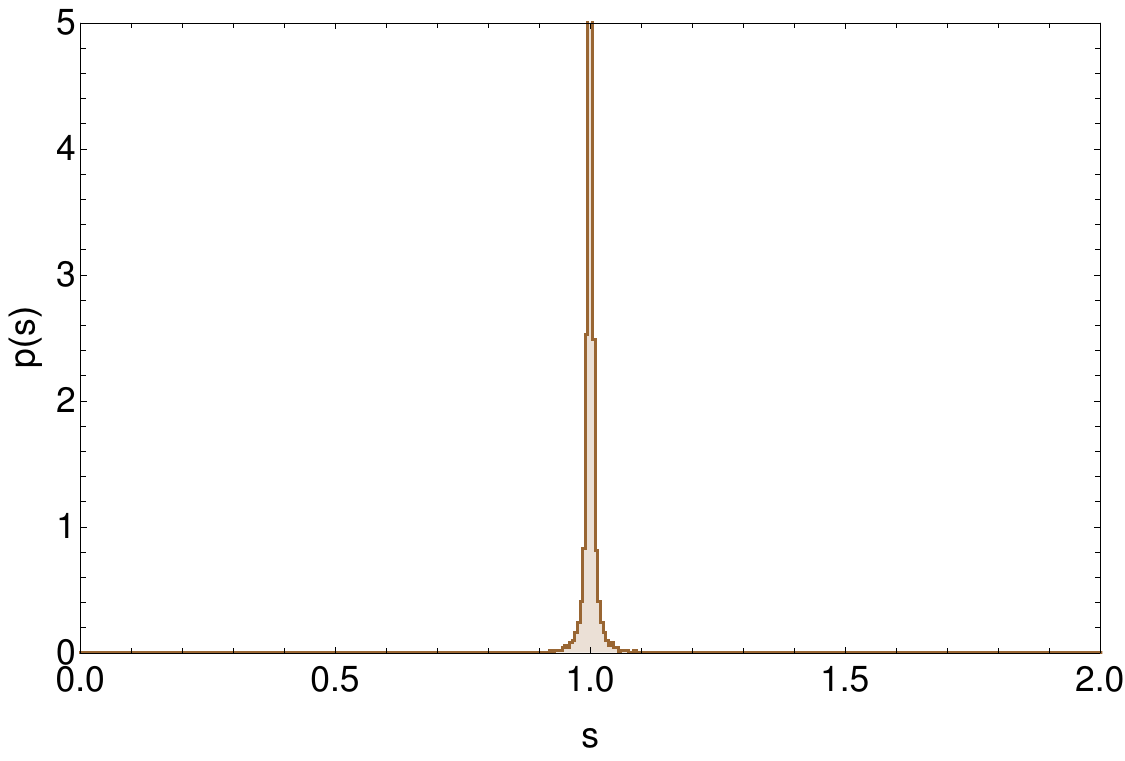}
    \end{subfigure}
    \hfill
    \begin{subfigure}{0.47\textwidth}
    \includegraphics[width=\textwidth]{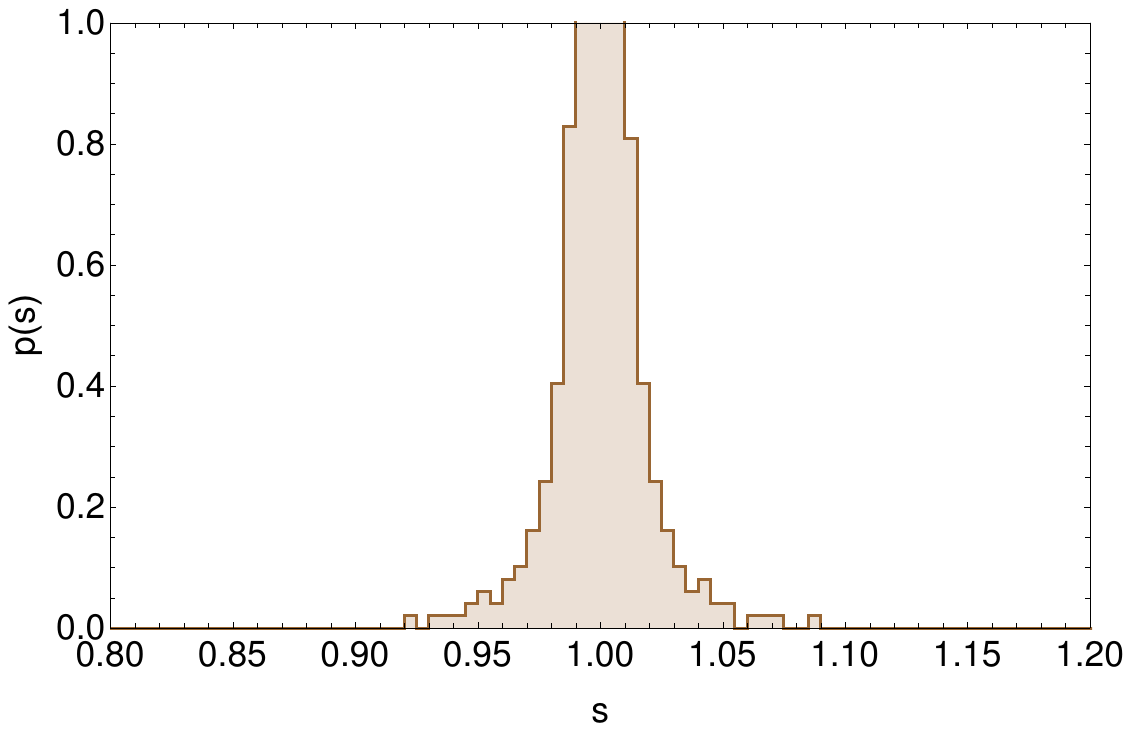}
    \end{subfigure}
    \caption{ LSD of log spectrum with $n_{\text{cut}}=10000$. There is no conventional level repulsion, but there is something akin to the ``extreme' level repulsion discussed in \cite{fuzz1,fuzz2}. The right panel shows the LSD zoomed in to show some structure.}
    \label{lsd-log}
\end{figure}

In the examples above, we have worked with fully deterministic spectra and we observed that it is possible for them to have ramps. When the ramp is linear, it is often viewed as a signature of RMT behavior and our calculations show that this observation has some interesting wrinkles. In fact, it is easy to check (and indeed intuitively plausible) that spectra like the power law and log examples we considered do not exhibit conventional level repulsion. At best, they can be viewed as exhibiting an ``extreme" form of level repulsion as suggested in \cite{fuzz1, fuzz2}, see Figure \ref{lsd-log}.

In this section, we will present a simple mechanism (inspired by the observations in \cite{fuzz2}) for introducing level repulsion reminiscent of Wigner-Dyson, into deterministic spectra. The idea is to add a small noise correction to each energy level of the spectrum with mean zero and a variance that is small compared to the level spacing around $n$:
\bea
E_n^{new} = E_n + {\rm noise}. \label{noise}
\eea
In \cite{fuzz2} it was observed that a non-trivial random profile at the stretched horizon for a scalar field on the black hole background, leads to a small noise in the spectrum of normal modes. This lead to level repulsion that could be matched with the Wigner-Dyson distributions. Here we abstract away the observation that noise can generate level repulsion, and turn it into a general mechanism for modifying deterministic spectra. 

\begin{figure}[H]
\begin{subfigure}{0.47\textwidth}
    \centering
    \includegraphics[width=\textwidth]{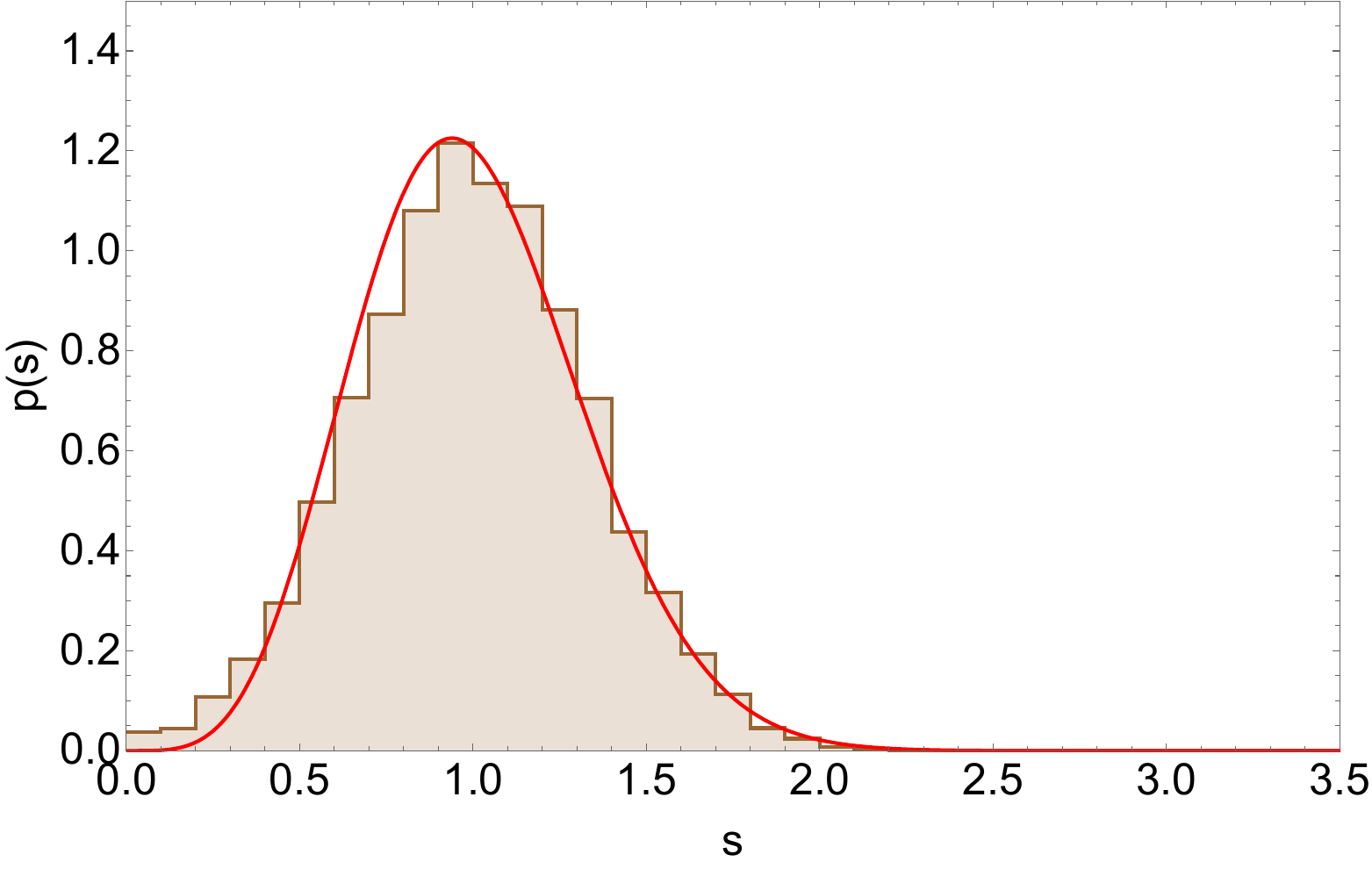}
    \end{subfigure}
    \hfill
    \begin{subfigure}{0.47\textwidth}
    \includegraphics[width=\textwidth]{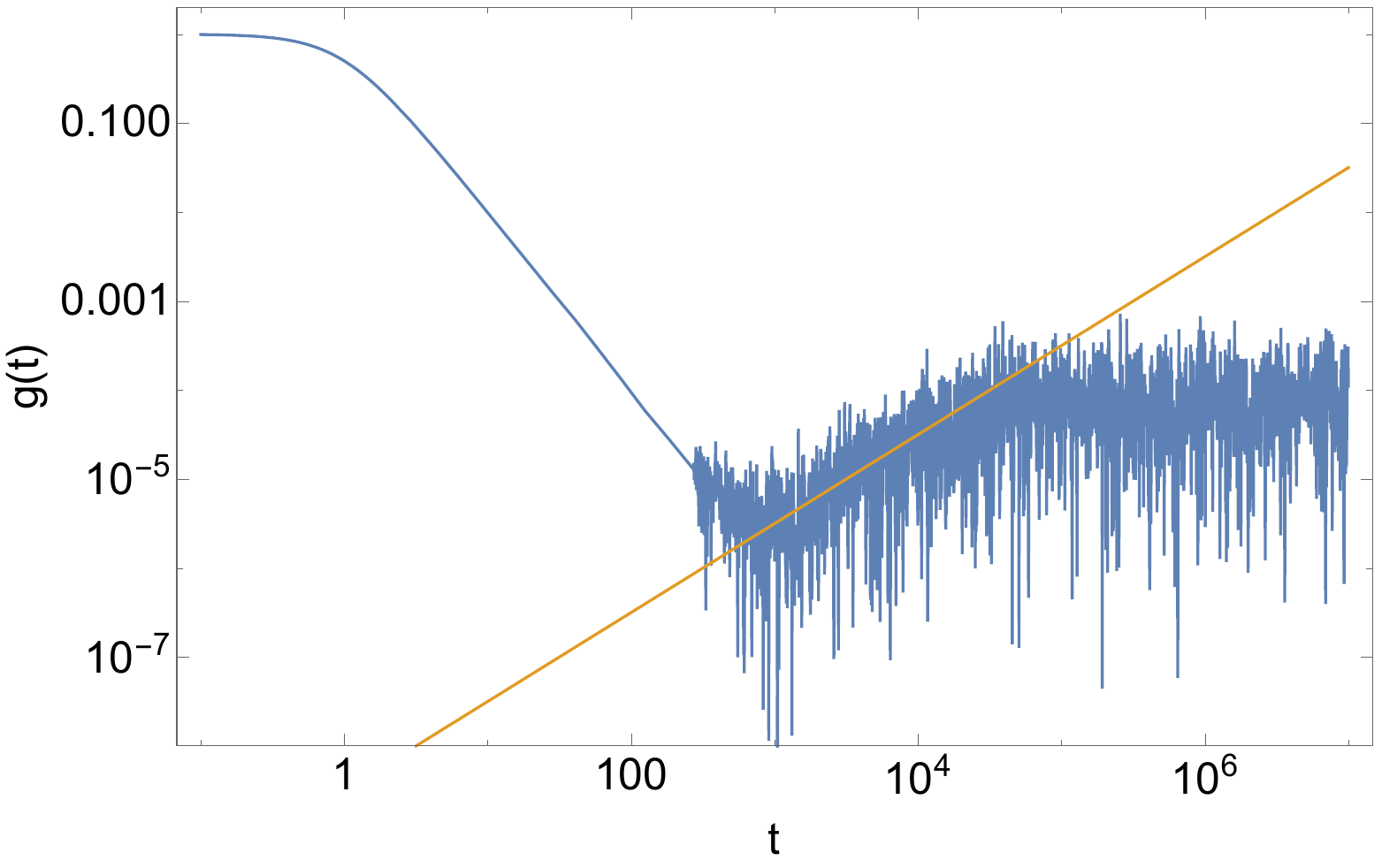}
    \end{subfigure}
    \caption{ LSD and SFF of the same noisy log spectrum with $n_{\text{cut}}=10000$. The red line is a Wigner-Dyson fit, and the variance of the noise at level $n$ has been chosen to be proportional to the level-spacing at $n$. The precise choices are unimportant. Slope of the yellow line is $1$. }
    \label{noisy_log}
\end{figure}

It is important to emphasize that we can introduce level repulsion of this kind, without sacrificing the slope of the ramp. We present the example of noisy log spectrum that exhibits level repulsion (while retaining the slope of the ramp in the noise-less counterpart) in Figure \ref{noisy_log}. 
Adding noise moves us away from the regime of deterministic sequences, so we will not discuss them much here. But it is worth emphasizing that the notion of randomness here is clearly quite different from the usual notion of RMT behavior. Yet, adding noise to certain deterministic spectra (like the log spectrum) results in both the linear ramp as well as level repulsion which are viewed as a random matrix diagnostics.

A further interesting use of adding a small amount of noise to deterministic ramps is that in some cases where there is no ramp, this can lead to the generation of a ramp (but often with slope $\neq 1$). This was first observed in \cite{fuzz2} where it was noted that adding noise leads to a ramp (albeit of slope $ \sim 1.9$) in the SHO spectrum, i.e., when $\alpha =1 $. We have checked that similar statement holds also  for some $\alpha >1$ cases, with appropriate choices of noise. The phenomenon seems quite generic and worth understanding better -- there is a possibility that since the spectra are deterministic this setting may be amenable to an analytic attack.
%
%
%
%

Finally, let us make one more comment about obtaining level repulsion, but {\em without} noise -- i.e., in an entirely deterministic way. This seems to be possible if instead of the noise term in \eqref{noise}, we add another function of $n$ that is suitably chosen. This second function can be (e.g.) quasi-periodic with an amplitude that is chosen to be a suitable function of $n$, and sufficiently smaller than $E_{n+1}-E_n$ so that it is effectively adding a variance to $E_n$. With suitable choices, this function's effect on the LSD can approximately resemble the level repulsion due to a noise term -- we have checked this. Of course, reproducing a Wigner-Dyson spectrum closely may require fine-tuning and post-selecting these functions considerably. The functions we found which could produce some amount of level repulsion are contrived and unlikely to be of physical interest, so we will not pursue this possibility here. But it is worth noting that both ramps and level repulsion seem possible, in fully deterministic sequences.

\section{Discussion}

A quantum notion of chaos is both ubiquitous and elusive. In order to characterize late time quantum chaos, the linearity of the ramp in the SFF as well as level repulsion in spectral statistics are often used. The lore is that these two features arise concurrently in spectra\footnote{We thank Julian Sonner for discussions on this point. Note that ensemble averages in integrable spectra can give rise to ramp-like structures \cite{Jeff}, but they are certainly not linear and probably not even power law \cite{fuzz1}. Note also that here, we do not need to  do any ensemble averaging.}. In previous papers \cite{fuzz1, fuzz2} we noticed that both the linear ramp as well as level repulsion can arise in black hole normal modes. But the linearity of the ramp was linked to the closeness of the stretched horizon to the actual horizon, while level repulsion was associated to the (random\footnote{Let us emphasize that the many notions of randomness should not be conflated. (Gaussian) randomness in the boundary condition profile leads to (Wigner-Dyson-like) randomness in the spectrum. Note that (for example) simply demanding a Gaussian distributed set of eigenvalues in a spectrum would {\em not} lead to level repulsion. So mere randomness of {\em any} sort, is not equivalent to random matrix behavior. Our motivation for trying a generic boundary profile, was the fact that BPS fuzzball microstates have generic profiles at the cap. Remarkably, this lead to level repulsion in the spectrum.}) choice of the profile function at the stretched horizon. The fact that these two features could be associated to two distinct features of the construction suggested that it may be possible to arrange this more generally away from the context of black holes. 

In this paper, we have explored this systematically by working with deterministic spectra. We provided sufficiency conditions for ramps to arise in power law spectra and noticed that the log spectrum is a tantalizingly simple example of a spectrum with a linear ramp. We also presented some analytic criteria that may be useful more generally, when searching for ramps in deterministic sequences. The connection between the log spectrum and the Riemann zeta function was used to make some connections between the linearity of the ramp and known theorems/conjectures about the Riemann zeta function. The fact that the ramp of zeta function is eternal and Platonic enabled us to compute its slope fairly effortlessly and show that it is numerically very close to 1 (i.e., to more decimals than is usually possible for typical systems like SYK or random matrices \cite{SYK}). It was further noted that mildly noisy spectra have level repulsion, giving us a systematic mechanism for engineering both the (linear) ramp and level repulsion simultaneously. This provides us a different path that is not based on random matrices to generate the linear ramp and level repulsion together.

These observations raise various questions about the precise nature of quantum chaos. They show that neither the linear ramp nor level repulsion need be tied to random matrices directly\footnote{What one means by a ``random matrix" is a matrix chosen from (say) a Gaussian unitary ensemble. What we mean by the statement that these spectra do not belong to a random matrix, is (strictly speaking) simply that these are highly non-generic elements of the ensemble.}. The fact that we stumbled on these observations by looking for signatures of chaos in black hole normal
modes suggests that perhaps the ramp and level repulsion have $independent$ significance for quantum chaos, and not merely through their connection to random matrices. It will be very interesting to explore this further, a potentially relevant discussion can be found in \cite{Amit}. 

In recent work \cite{Pradipta} (to appear) we have found that the modes responsible for black hole entropy in 't Hooft's brickwall calculation \cite{tHooft} are indeed the ones responsible for the linear ramp.  Let us make some further comments about the log spectrum and its connection to black holes. In the works in \cite{fuzz1, fuzz2} we studied the normal mode spectrum numerically. In \cite{Pradipta}, an analytic expression 
\bea
E_J \sim \frac{1}{a-b \log J}
\eea
is provided for the low-lying part of the spectrum. (The details of the constants $a$ and $b$ are not important for our present discussion.). For relatively low values of $J$ (i.e., the lowest lying part of the spectrum) this reduces to 
\bea
E_J \sim \log J
\eea
after ignoring overall scales and shifts. In light of our results in this paper, this fact seems (at least partly) responsible for the origin of the linear ramp noted in \cite{fuzz1, fuzz2}. 

Continuing in this vain -- It is well-known \cite{Julia} that the log spectrum arises in a Fock space constructed from multi-particling a bosonic Hilbert space where the orthonormal states of the single particle Hilbert space are $|p\rangle$ (labeled by primes $p$) with energy $\log p$. This system is called the primon gas \cite{Primon}. Unique factorization of integers $n$ over primes then leads to the $\log n$ spectrum and the Riemann zeta function emerges as the (complexified) partition function of the primon gas. In the context of supersymmetric black holes, various number theoretic partition functions have played a significant role in string theory. It is therefore tantalizing that the Riemann zeta function has emerged in the present discussion for generic (finite temperature) black holes, via the low lying $\log n$ behavior of normal mode spectra. Riemann zeta function as a partition function, naturally leads to the interpretation that the pole at $s=1$ is related to a Hagedorn transition. In the context of black holes \cite{Pradipta}, this scale is controlled by the stretched horizon size which has sometimes been connected to the string scale \cite{Mertens}. It will be very interesting to connect the analytic continuation of the RZF to $\beta=0$ via a string-black hole transition \cite{Susskind, HP}. From this perspective, perhaps it is natural then that the RZF obtained this way is a ``more classical" object with $n_{cut}=\infty$ as demonstrated by the eternal ramp.


\section{Acknowledgments}

CK thanks Rajesh Gopakumar and Vyshnav Mohan for discussions/correspondence and the University of the Witwatersrand for hospitality during the final stages of this work. AK is partially supported by CEFIPRA 6304 -- 3, DAE-BRNS 58/14/12/2021-BRNS and CRG/2021/004539 of Govt. of India. AK further thanks the warm hospitality of the Department of Theoretical Physics, CERN during the completion of this work. The work of SKG has been supported by SERB, DST, India through
grant TAR/2023/000116.

\appendix

\section{Log Potential For Log Spectrum} 

As we noted, the $\log$ spectrum is closely related to the spectrum of black hole normal modes and is in that sense of immediate physical interest. But one may also seek a simple quantum mechanical 0+1 dimensional system whose spectrum $E_n \sim \log n$. In this section, we show that a $\log$ spectrum is closely related to a $\log q$  potential. Modulo the fact that it needs some modification  around $q=0$ for bounding it from below, this provides a potential with the requisite spectrum. 

Given a spectrum, the inverse problem of constructing the corresponding potential may not be unique and is therefore not mathematically always well-defined. Nonetheless, within a WKB description, it is sometimes possible to solve this inverse problem for classes of spectra (sometimes with mild extra assumptions). Motivated by this, here we will explicitly construct a potential that has a logarithmic semi-classical spectrum. The results here are undoubtedly well-known, we present them only to demonstrate that the $\log n$ spectrum is not particularly exotic. 

Let us begin with the Bohr-Sommerfeld quantization condition:
\begin{eqnarray}
2 \pi n = \oint p dq \ ,
\end{eqnarray}
where we have set $\hbar=1$, $p,q$ are the classical momenta and co-ordinates and $n\in {\mathbb Z}$. Assuming a non-relativistic system, the RHS of the above expression can be re-written in terms of the energy and the potential of the corresponding Hamiltonian, which yields:
\begin{eqnarray}
2 \pi n & = & \oint \sqrt{2m \left( E - V(q) \right)} dq  \nonumber\\
& = & 2 \int_0^{q_{\rm max}} \sqrt{2m \left( E - V(q) \right)} dq \ ,
\end{eqnarray}
where, in the last step, we have assumed that $V(q)$ is symmetric around the $q=0$ point and $V(q_{\rm max}) =E$. Taking a derivative with respect to $E$ on both sides above, we get:
\begin{eqnarray}
\pi \frac{\partial n}{\partial E} = \int_0^{q_{\rm max}} \frac{\sqrt{2 m} }{\sqrt{E - V(q)}} dq .
\end{eqnarray}
To invert the integral formula above, we multiply both sides by $(2m(x-E))^{-1/2}$ and integrate both sides with respect to $E$. This yields:
\begin{eqnarray}
\int_0^{x} \frac{\partial n}{\partial E} \frac{dE}{\sqrt{2m(x - E)}} = q_{\rm max} = V^{-1}(x) \ . 
\end{eqnarray}
The last equation follows from the fact that $x$ scales like $E$. Thus, if we are able to evaluate the LHS, we can infer the potential.

Let us now consider the logarithmic spectrum: $E_n = \beta \log n$. This yields:
\begin{eqnarray}
&& \int_0^x \frac{e^{E/\beta}}{\beta} \frac{dE}{\sqrt{2m(x-E)}} = q_{\rm max} = V^{-1}(x) \ . \\
&& \implies q_{\rm max} = - e^{x/\beta} \sqrt{\frac{\pi}{2m\beta}} \text{erfc}\left(\sqrt{\frac{x}{\beta }}\right) \ .
\end{eqnarray}
To obtain $V(q)$, we need to invert the above relation to obtain $x(q_{\rm max})$. An exact analytical inversion may not be possible, however, in the limit $x \ll 1$ and $x \gg 1$, this relation can be inverted to obtain:
\begin{eqnarray}
&& q_{\rm max} = - \sqrt{\frac{\pi}{2m\beta}} e^{x/\beta} + {\cal O}(x) \ , x\ll 1 \ , \\
&& q_{\rm max} = - \sqrt{\frac{1}{2m x}} + {\cal O}(x^{-3/2}) \ , x\gg 1 \ ,
\end{eqnarray}
In the two respective regimes, the above inversion formulae yield the following behaviour of the potential:
\begin{eqnarray}
&& V(q) \sim \log q \ , \quad q \sim 1 \ , \\
&& V(q) \sim \frac{1}{q^2} \ , \quad q \ll 1 \ .
\end{eqnarray}
In writing the above expressions, we have ignored constant terms in the potential, as well as various factors where various parameters of the spectrum are also present. This exercise explicitly demonstrates that a logarithmic spectrum can emerge from a potential which has a logarithmic behaviour with some minor modifications to make it bounded below.


\end{document}